\documentclass[12pt]{article}\usepackage[]{graphicx}\usepackage[]{color}
\makeatletter
\def\maxwidth{ %
  \ifdim\Gin@nat@width>\linewidth
    \linewidth
  \else
    \Gin@nat@width
  \fi
}
\makeatother

\definecolor{fgcolor}{rgb}{0.345, 0.345, 0.345}

\usepackage{framed}
\makeatletter
 {\par\unskip\endMakeFramed%
 \at@end@of@kframe}
\makeatother

\definecolor{shadecolor}{rgb}{.97, .97, .97}
\definecolor{messagecolor}{rgb}{0, 0, 0}
\definecolor{warningcolor}{rgb}{1, 0, 1}
\definecolor{errorcolor}{rgb}{1, 0, 0}
\usepackage{alltt}
\usepackage[latin9]{inputenc}
\usepackage{geometry}
\usepackage{amsmath}
\usepackage{amssymb}
\usepackage{setspace}
\usepackage{natbib}
\setcitestyle{aysep={}}
\usepackage[labelfont=bf,labelsep=period,font={small,sl}]{caption}
\usepackage{amsthm}
\usepackage{epsfig}
\usepackage{psfrag}
\usepackage{booktabs}
\usepackage{lineno}
\usepackage{url}

\renewcommand{\baselinestretch}{1.3}
\IfFileExists{upquote.sty}{\usepackage{upquote}}{}
\begin{document}

%\VignetteIndexEntry{Guide to using momentuHMM}
%\VignetteEngine{knitr::knitr}

% set margin to 4cm for title page
\newgeometry{margin=4cm}

\begin{center}
  \texttt{\LARGE momentuHMM}{\LARGE : R package for generalized hidden Markov models of animal movement}\vspace{0.5in}
  \par
\end{center}

\begin{center}
  {\large Brett T. McClintock$^{1}$ and Th\'eo Michelot$^{2}$} 
  \par
\end{center}

\begin{center}
  \hrulefill{} 
  \par
\end{center}

\begin{center}
  \global\long\def\baselinestretch{1.25}
  {\large $^{1}$Marine Mammal Laboratory}\\
  {\large Alaska Fisheries Science Center}\\
  {\large {} NOAA National Marine Fisheries Service}\\
  {\large {} Seattle, U.S.A.}\\
  {\large {} {\em Email:} brett.mcclintock@noaa.gov} 
  \par
\end{center}

{\large \par}

\begin{center}
  {\large $^{2}$School of Mathematics and Statistics}\\
  {\large University of Sheffield}\\
  {\large {} Sheffield, U.K.}
  \par
\end{center}

\begin{center}
  {\large \hrulefill{}} 
  \par
\end{center}

\begin{center}
  \textsc{Running Head}: R package \verb|momentuHMM| \bigskip{}  
  \par
\end{center}

\begin{center}
  \today
  \par
\end{center}

\clearpage{}

% \setlength{\textheight}{575pt} \global\long\def\baselinestretch{2}
%% ABSTRACT %%%%%%%%%%%%%%%%%%%%%%%%%%%%%%%%%%%%%%%%%%%%%%%%%%%%%

%  make sure that the document has 25 lines per page (it is 12 pt)
% \setlength{\textheight}{575pt} \setlength{\baselineskip}{24pt} %think this may do doublespacing (required for submission I think)

\newpage{}

%\linenumbers

% set margin to 3cm for main text
\newgeometry{margin=3cm}

\noindent \textbf{Summary}\\
\textbf{1.} Discrete-time hidden Markov models (HMMs) have become an immensely popular tool for inferring latent animal behaviours from telemetry data. While movement HMMs typically rely solely on location data (e.g.\ step length and turning angle), auxiliary biotelemetry and environmental data are powerful and readily-available resources for incorporating much more ecological and behavioural realism. However, complex movement or observation process models often necessitate custom and computationally-demanding HMM model-fitting techniques that are impractical for most practitioners, and there is a paucity of generalized user-friendly software available for implementing multivariate HMMs of animal movement.\\
\textbf{2.} Here we introduce an open-source R package, \verb|momentuHMM|, that addresses many of the deficiencies in existing HMM software.  Features include: 1) data pre-processing and visualization; 2) user-specified probability distributions for an unlimited number of data streams and latent behaviour states; 3) biased and correlated random walk movement models, including dynamic ``activity centres'' associated with attractive or repulsive forces; 4) user-specified design matrices and constraints for covariate modelling of parameters using formulas familiar to most R users; 5) multiple imputation methods that account for measurement error and temporally-irregular or missing data; 6) seamless integration of spatio-temporal covariate raster data; 7) cosinor and spline models for cyclical and other complicated patterns; 8) model checking and selection; and 9) simulation.\\
\textbf{3.} After providing an overview of the main features of the package, we demonstrate some of the capabilities of \verb|momentuHMM| using real-world examples. These include models for cyclical movement patterns of African elephants, foraging trips of northern fur seals, loggerhead turtle movements relative to ocean surface currents, and grey seal movements among three activity centres.\\
\textbf{4.} \verb|momentuHMM| considerably extends the capabilities of existing HMM software while accounting for common challenges associated with telemetry data. It therefore facilitates more realistic hypothesis-driven animal movement analyses that have hitherto been largely inaccessible to non-statisticians.  While motivated by telemetry data, the package can be used for analyzing any type of data that is amenable to HMMs. Practitioners interested in additional features are encouraged to contact the authors.\\
\noindent \textbf{Key-words} biologging, biotelemetry, \verb|crawl|, \verb|moveHMM|, state-space model, state-switching

\global\long\def\baselinestretch{1.0}
 \global\long\def\baselinestretch{1.0}

\section{Introduction}

Animal movement is central to many ecological processes and considered essential to our understanding of animal behaviour, population dynamics, and the impacts of global change. Coupled with advances in animal-borne biologging technology, there has been a recent explosion in analytical methods for inferring animal movement, behaviour, space use, and resource selection from telemetry data \citep{HootenEtAl2017}. Largely attributable to their ease of implementation and interpretation, discrete-time hidden Markov models (HMMs) have become immensely popular for characterizing animal movement behaviour \citep[e.g.][]{MoralesEtAl2004,JonsenEtAl2005,LangrockEtAl2012,McClintockEtAl2012}. 

In short, an HMM is a time series model composed of a (possibly multivariate) observation process $({\mathbf Z}_1,\ldots,{\mathbf Z}_T)$, in which each data stream is generated by $N$ state-dependent probability distributions, and where the unobservable (hidden) state sequence $(S_t\in\{1,\ldots,N\},t=1,\ldots,T)$ is assumed to be a Markov chain.  The state sequence of the Markov chain is governed by (typically first-order) state transition probabilities, $\gamma_{ij}^{(t)}=\text{Pr}(S_{t+1}=j \mid S_t=i)$ for $i,j=1,\ldots,N$, and an initial distribution ${\boldsymbol \delta}^{(0)}$.  The likelihood of an HMM can be succinctly expressed using the forward algorithm:
\begin{equation}
  \mathcal{L}={\boldsymbol \delta}^{(0)} {\mathbf \Gamma}^{(1)} {\mathbf P}({\mathbf z}_1) {\mathbf \Gamma}^{(2)} {\mathbf P}({\mathbf z}_2) {\mathbf \Gamma}^{(3)} \cdots {\mathbf \Gamma}^{(T-1)} {\mathbf P}({\mathbf z}_{T-1}) {\mathbf \Gamma}^{(T)} {\mathbf P}({\mathbf z}_{T}) {\bf 1}^N,
  \label{eq:HMMlike}
\end{equation}
where ${\mathbf \Gamma}^{(t)}=\left(\gamma_{ij}^{(t)} \right)$ is the $N \times N$ transition probability matrix, ${\mathbf P}({\mathbf z}_t)=\text{diag}(p_1({\mathbf z}_t), \ldots, p_N({\mathbf z}_t))$, $p_s({\mathbf z}_t)$ is the conditional probability density of ${\mathbf Z}_t$ given $S_t=s$, and ${\bf 1}^N$ is a $N$-vector of ones. For a thorough introduction to HMMs, see \cite{ZucchiniEtAl2016}.  

The most common discrete-time animal movement HMMs for telemetry location data are composed of two data streams (e.g. step length and turning angle) and $N=2$ hidden states \citep{MoralesEtAl2004,JonsenEtAl2005}. These states are often considered as proxies for animal behaviour and characterized by area-restricted-search-type movements  (``encamped'' or ``foraging'') and migratory-type movements (``exploratory'' or ``transit''). Some common probability distributions for the step length data stream are the gamma or Weibull distributions, while the wrapped Cauchy or von Mises distributions are often employed for the turning angle \citep[e.g.][]{LangrockEtAl2012}. While HMMs based solely on location data are somewhat limited in the number and type of biologically-meaningful movement behaviour states they are able to accurately identify, auxiliary biotelemetry and environmental data allow for multivariate HMMs that can incorporate much more behavioural realism and facilitate inferences about complex ecological relationships that would otherwise be difficult or impossible to infer from location data alone \citep[e.g.][]{McClintockEtAl2013c,McClintockEtAl2017,DeRuiterEtAl2017}.  

When data streams are observed without error and at regular time intervals, a major advantage of HMMs is the relatively fast and efficient maximization of the likelihood using the forward algorithm (Eq.\ \ref{eq:HMMlike}) and computation of the most likely sequence of hidden states using the Viterbi algorithm \citep{ZucchiniEtAl2016}. In movement HMMs, the decoded state sequence can be useful for determining when and where changes in behaviour occur, identifying habitats associated with particular behaviours (e.g. foraging ``hotspots''), and calculating the proportion of time steps allocated to each state (i.e. ``activity budgets''). However, location measurement error is rarely non-existent in animal-borne telemetry studies and depends on both the device and the system under study \citep[e.g.][]{CostaEtAl2010}. Furthermore, telemetry data are often obtained with little or no temporal regularity, as in many marine mammal telemetry studies \citep[e.g.][]{JonsenEtAl2005}, such that observations do not align with the regular time steps required by discrete-time HMMs. When explicitly accounting for uncertainty attributable to location measurement error, temporally-irregular observations, or other forms of missing data, one must typically fit HMMs using computationally-intensive (and often time-consuming) model fitting techniques such as Markov chain Monte Carlo \citep[e.g.][]{JonsenEtAl2005,McClintockEtAl2012}. Unfortunately, complex analyses requiring novel statistical methods and custom model-fitting algorithms are not practical for many practitioners.

While statisticians have been applying HMMs to telemetry data for decades, R \citep{RCoreTeam2017} packages such as \verb|bsam| \citep{JonsenEtAl2005}, \verb|moveHMM| \citep{MichelotEtAl2016}, and \verb|swim| \citep{WhoriskeyEtAl2017} have recently helped make these models more accessible to the practitioners that are actually conducting telemetry studies. While these contributions represent important steps %toward making HMMs of animal movement more accessible
forward, the models that can currently be implemented remain limited in many key respects. For example, most HMM software for animal movement is limited to correlated random walks with $N=2$ states and two data streams (e.g.\ step length and turning angle). While \verb|moveHMM| does permit $N>2$, it is typically difficult to identify $>$2 biologically-meaningful behaviour states from only two data streams \citep[e.g.][]{MoralesEtAl2004,BeyerEtAl2013,McClintockEtAl2014b}. Both \verb|moveHMM| and \verb|swim| require temporally-regular location data with negligible measurement error%, but the realities of animal-borne telemetry often yield temporally-irregular location data subject to error (particularly in aquatic environments)
. Other notable deficiencies of existing software include limited abilities to incorporate spatio-temporal environmental or individual covariates on parameters, biased movements in response to attractive, repulsive, or environmental forces \citep[e.g.][]{McClintockEtAl2012}, group dynamic models \citep{LangrockEtAl2014}, cyclical (e.g.\ daily, seasonal) and other more complicated behavioural patterns, or constraints on parameters. %For temporally-regular data with no measurement error, the univariate HMM R package \verb|HiddenMarkov| \citep{Harte2017} is very flexible and has recently been extended to accomodate covariates on the mean parameter of a single data stream, but it can not do so with multivariate data or probability distributions commonly used in movement HMMs (e.g.\ Weibull, von Mises, wrapped Cauchy).

To address these deficiencies in existing software, we introduce a new user-friendly R package, \verb|momentuHMM|% (Maximum likelihood analysis Of animal MovemENT behaviour Using multivariate Hidden Markov Models)
, intended for practitioners wishing to implement more flexible and realistic HMM analyses of animal movement while accounting for common challenges associated with telemetry data. %Features for multivariate HMM analyses in \verb|momentuHMM| include: 1) tools for data pre-processing and visualization; 2) user-specified probability distributions for an unlimited number of data streams and latent behaviour states; 3) user-specified design matrices and constraints for covariate modelling of state transition probability and probability distribution parameters using linear model formulas familiar to most R users; 4) multiple imputation methods that account for observation error attributable to measurement error and temporally-irregular or missing data \citep{HootenEtAl2017,McClintock2017}; 5) seamless integration of spatio-temporal environmental covariate data (e.g.\ wind direction, forest cover, sea ice concentration) using the \verb|raster| package \citep{Hijmans2016}; 6) incorporation of ``activity centres'' associated with attractive or repulsive forces \citep[e.g.][]{McClintockEtAl2012,MichelotEtAl2017}; 7) circular-circular regression models for angular probability distributions \citep{DuchesneEtAl2015}; 8) cosinor \citep[e.g.][]{Cornelissen2014} and spline regression formulas for cyclical and other complicated behavioural patterns; and 9) data simulation capabilities for power analyses and assessing model performance, including simulation of location data subject to temporal irregularity and/or measurement error.  
We provide a brief overview of \verb|momentuHMM| and demonstrate some of its capabilities using real-world examples. %These include an example of periodic movement cycles in African elephants, a 3-state (``resting'', ``foraging'', ``transit'') northern fur seal example incorporating auxiliary dive activity data \citep{McClintockEtAl2014b}, a loggerhead turtle example relating ``foraging'' and ``transit'' movements to ocean surface currents, a 5-state grey seal example incorporating biased movements toward haul-out and foraging locations \citep{McClintockEtAl2012}, and a 4-state (``outbound'', ``searching'', ``foraging'', ``inbound'') southern elephant seal example with biased movements toward and away from a colony \citep{MichelotEtAl2017}. %While many of the features of \verb|momentuHMM| were motivated by animal movement data, the package can be used for analyzing any type of data that is amenable to (multivariate) HMMs.  
Step-by-step tutorials, help files, examples, and code can be found in the package's documentation and vignette. This article describes \verb|momentuHMM| version 1.4.0.

\section{momentuHMM overview}
\begin{figure}[htbp]
  \centering
  \includegraphics[width=\textwidth]{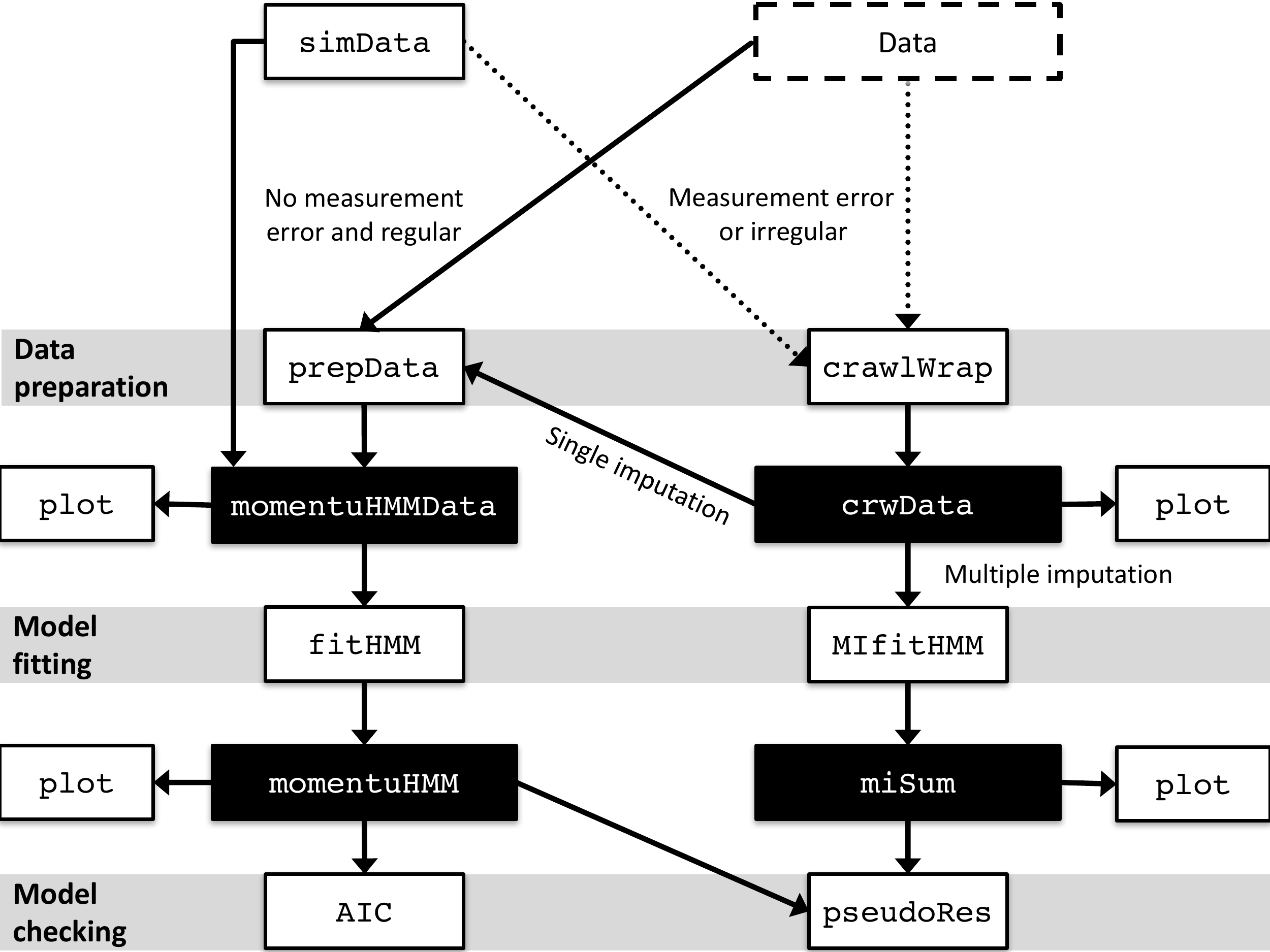}
  \caption{Schematic representing the typical momentuHMM workflow. White boxes indicate package functions and black boxes indicate object classes returned by functions. The simData function can be used to simulate multivariate HMM data from scratch or from a fitted model. When telemetry data are subject to location measurement error or temporal irregularity (dotted arrows), these sources of uncertainty can be incorporated into HMMs using single or multiple imputation of the position process via the crawlWrap function.}
  \label{fig:workflow}
\end{figure}

%We first provide an overview of the main features and functions of \verb|momentuHMM|.  
The workflow for \verb|momentuHMM| analyses consists of several steps (Fig.\ \ref{fig:workflow}). These include data preparation, data visualization, model specification and fitting, results visualization, and model checking. The workhorse functions of the package are listed in Table \ref{tab:functions}. Usage of several of these functions is deliberately similar to equivalent functions in \verb|moveHMM| \citep{MichelotEtAl2016}, but the \verb|momentuHMM| arguments have been generalized and expanded to accommodate a much more flexible framework for data pre-processing, model specification, parameterization, visualization, and simulation. R users already familiar with \verb|moveHMM| will therefore likely find it easy to immediately begin using \verb|momentuHMM|. %While the \verb|momentuHMM| syntax is therefore more complicated, any \verb|moveHMM| model can be implemented in \verb|momentuHMM|.

\begin{table}
  \caption{\label{tab:functions} Workhorse functions for the R package momentuHMM.}
  \begin{tabular}{ll}
  \toprule
  Function & Description \tabularnewline
  \midrule
  %\verb|AIC.momentuHMM| & AIC for one or several \verb|momentuHMM| models  \tabularnewline
  %\verb|CIbeta| & Confidence intervals for working (beta) parameters  \tabularnewline
  %\verb|CIreal| & Confidence intervals for natural (real) parameters  \tabularnewline
  \verb|crawlMerge| & Merge \verb|crawlWrap| output with additional data streams or covariates  \tabularnewline 
  \verb|crawlWrap| & Fit continuous-time correlated random walk (CTCRW) models and \\ &predict temporally-regular locations using the \verb|crawl| package \tabularnewline  
  \verb|fitHMM| & Fit a (multivariate) HMM to the data \tabularnewline  
  \verb|MIfitHMM| & Fit (multivariate) HMMs to multiple imputation data  \tabularnewline  
  \verb|MIpool| & Pool \verb|momentuHMM| model results across multiple imputations  \tabularnewline 
  \verb|plot.crwData| & Plot \verb|crawlWrap| output \tabularnewline 
  \verb|plot.miSum| & Plot summaries of multiple imputation \verb|momentuHMM| models  \tabularnewline 
  \verb|plot.momentuHMM| & Plot summaries of \verb|momentuHMM| models  \tabularnewline 
  \verb|plot.momentuHMMData| & Plot summaries of selected data streams and covariates  \tabularnewline 
  \verb|plotPR| & Plot time series, qq-plots and sample ACFs of pseudo-residuals \tabularnewline 
  \verb|plotSat| & Plot locations on satellite image \tabularnewline   
  \verb|plotSpatialCov| & Plot locations on raster image \tabularnewline   
  \verb|plotStates| & Plot the (Viterbi-)decoded states and state probabilities \tabularnewline 
  \verb|prepData| & Pre-process data streams and covariates \tabularnewline 
  \verb|pseudoRes| & Calculate pseudo-residuals for \verb|momentuHMM| models \tabularnewline 
  \verb|simData| & Simulate data from a (multivariate) HMM \tabularnewline 
  \verb|stateProbs| & State probabilities for each time step \tabularnewline 
  \verb|viterbi| & Most likely state sequence (using the Viterbi algorithm)  \tabularnewline  
  \bottomrule
  \end{tabular}
\end{table}

A key feature of \verb|momentuHMM| is the ability to include an unlimited number of HMM data streams (e.g.\ step length, turning angle, dive activity, heart rate) arising from a broad range of commonly used probability distributions (e.g.\ beta, gamma, normal, Poisson, von Mises, Weibull). Using model formulas familiar to most R users, any of the data stream probability distribution parameters can be modelled as functions of environmental and individual covariates via link functions \citep[e.g.][and see Table \ref{tab:pdfs}]{McCullaghNelder1989}. %For any given ``natural scale'' (or ``real scale'') probability distribution parameter $\theta$, all of the link functions $(g)$ in \verb|momentuHMM| are of the general form $g({\boldsymbol \theta}) =  {\mathbf X}_\theta{\boldsymbol \beta}_\theta$, where ${\mathbf X}_\theta$ is the $T \times k$ design matrix (composed of $k$ covariates) and ${\boldsymbol \beta}_\theta$ is the correponding $k$-vector of ``working scale'' (or ``beta scale'') parameters for $\theta$. For example, suppose step length is assumed to have a gamma distribution, $l_t\mid S_t=s \sim \text{gamma}(\mu_s,\sigma_s)$. In \verb|momentuHMM|, the natural scale parameters for the gamma distribution are the (state-dependent) step length mean $(\mu_s>0)$ and standard deviation $(\sigma_s>0)$.  Because both of these parameters must be positive, the log link function is a natural choice for modelling these parameters as a function of covariates, e.g.\ $\log({\boldsymbol \mu}) =  {\mathbf X}_\mu  {\boldsymbol \beta}_\mu$.  
The initial distribution $({\boldsymbol \delta}^{(0)})$ and state transition probabilities $(\gamma_{ij}^{(t)})$ can also be modelled as functions of covariates% using a multinomial logit link
. Permissable R classes for covariates include \verb|numeric|, \verb|integer|, or \verb|factor|. %Factors can be particularly useful for specifying models with individual- or group-level (e.g.\ sex or age class) effects on state transition and probability distribution parameters. 
Spatio-temporal covariates (e.g.\ wind velocity, forest cover, sea ice concentration) can also be \verb|raster| classes \citep{Hijmans2016}, in which case \verb|momentuHMM| extracts the appropriate values based on the time and location of each observation (see section \ref{sec:turtle}).

\begin{table}
  \caption{\label{tab:pdfs} Data stream $(z)$ probability distributions, natural parameters, and default link functions for covariate modelling. Probability distributions with positive support can be zero-inflated (with additional zero-mass parameters), while the beta distribution can be zero- and one-inflated (with additional one-mass parameters). If user-specified bounds are provided, then custom link functions are used instead of the defaults (see momentuHMM package documentation for further details). If circular-circular regression is specified for the mean of angular distributions (``vm'' and ``wrpcauchy''), then a link function based on \cite{RivestEtAl2016} is used. The von Mises consensus distribution (``vmConsensus'') is a von Mises circular-circular regression model where the concentration parameter depends on the level of agreement among short-term directional persistence and angular covariates. Users seeking additional data stream probability distributions are encouraged to contact the authors.}
  \begin{tabular}{llll}
  \toprule
  Distribution & Support & Parameters & Link function\footnotemark \tabularnewline
  \midrule
  Bernoulli (``\verb|bern|'')          & $z_t\in\{0,1\}$          & $\verb|prob|\in(0,1)$            &  $\text{logit}$ \tabularnewline  
  Beta (``\verb|beta|'')               & $z_t\in(0,1)$            & $\verb|shape1|>0$                &  $\log$ \tabularnewline  
                                       &                          & $\verb|shape2|>0$                &  $\log$ \tabularnewline
                                       &                          & $\verb|zero-mass|\in(0,1)$       &  $\text{logit}$ \tabularnewline 
                                       &                          & $\verb|one-mass|\in(0,1)$        &  $\text{logit}$ \tabularnewline 
  Exponential (``\verb|exp|'')         & $z_t>0$                  & $\verb|rate|>0$                  &  $\log$ \tabularnewline  
                                       &                          & $\verb|zero-mass|\in(0,1)$       &  $\text{logit}$ \tabularnewline 
  Gamma (``\verb|gamma|'')             & $z_t>0$                  & $\verb|mean|>0$                  &  $\log$ \tabularnewline  
                                       &                          & $\verb|sd|>0$                    &  $\log$ \tabularnewline  
                                       &                          & $\verb|zero-mass|\in(0,1)$       &  $\text{logit}$ \tabularnewline 
  Log normal (``\verb|lnorm|'')        & $z_t>0$                  & $\verb|location|\in{\rm I\!R}$   &  identity \tabularnewline  
                                       &                          & $\verb|scale|>0$                 &  $\log$ \tabularnewline  
                                       &                          & $\verb|zero-mass|\in(0,1)$       &  $\text{logit}$ \tabularnewline 
  Normal (``\verb|norm|'')             & $z_t\in{\rm I\!R}$       & $\verb|mean|\in{\rm I\!R}$       &  identity \tabularnewline  
                                       &                          & $\verb|sd|>0$                    &  $\log$ \tabularnewline 
  Poisson (``\verb|pois|'')            & $z_t\in\{0,1,\ldots\}$   & $\verb|lambda|>0$                &  $\log$ \tabularnewline  
  Von Mises (``\verb|vm|'')            & $z_t\in(-\pi,\pi]$       & $\verb|mean|\in(-\pi,\pi]$       &  $\tan(\verb|mean|/2)$ \tabularnewline  
                                       &                          & $\verb|concentration|>0$         &  $\log$ \tabularnewline 
  Von Mises (``\verb|vmConsensus|'')   & $z_t\in(-\pi,\pi]$       & $\verb|mean|\in(-\pi,\pi]$       &  \cite{RivestEtAl2016} \tabularnewline  
                                       &                          & $\verb|kappa|>0$                 &  $\log$ \tabularnewline 
  Weibull (``\verb|weibull|'')         & $z_t>0$                  & $\verb|shape|>0$                 &  $\log$ \tabularnewline  
                                       &                          & $\verb|scale|>0$                 &  $\log$ \tabularnewline  
                                       &                          & $\verb|zero-mass|\in(0,1)$       &  $\text{logit}$ \tabularnewline 
  Wrapped Cauchy (``\verb|wrpcauchy|'')& $z_t\in(-\pi,\pi]$       & $\verb|mean|\in(-\pi,\pi]$       &  $\tan(\verb|mean|/2)$ \tabularnewline  
                                       &                          & $\verb|concentration|\in(0,1)$ &  $\text{logit}$ \tabularnewline 
  \bottomrule
  \end{tabular}
  \footnotesize{$^1$Link functions $(g)$ relate natural scale parameters $({\boldsymbol \theta})$ to a $T \times k$ design matrix $({\mathbf X})$ and $k-$vector of working scale parameters $(\boldsymbol{\beta}\in \mathbb{R}^k)$ such that $g({\boldsymbol \theta})={\mathbf X}\boldsymbol{\beta}$.}
\end{table}

\subsection{Data preparation and visualization}
For temporally-regular location data with negligible measurement error, the \verb|prepData| function returns a \verb|momentuHMMData| object (i.e. a data frame of class \verb|momentuHMMData|) that can be used for data visualization and further analysis. Summary plots of \verb|momentuHMMData| objects can be created for any data stream or covariate using the generic \verb|plot| function. If location data are temporally-irregular or subject to measurement error, then a 2-stage multiple imputation approach can be performed \citep{HootenEtAl2017,McClintock2017}. We discuss this pragmatic approach to incorporating location uncertainty into HMMs of animal movement in section \ref{sec:mi}.

\subsection{HMM specification and fitting}
%Once a \verb|momentuHMMData| object has been created using \verb|prepData|, then the data are ready to be passed to the generalized multivariate HMM-fitting function \verb|fitHMM|. 
The function \verb|fitHMM| is used to specify and fit a (multivariate) HMM using maximum likelihood methods.  There are many different options for \verb|fitHMM|, so here we will only focus on several of the most important and useful features. The bare essentials of \verb|fitHMM| include the arguments:
\begin{itemize}
  \item{\verb|data|} A \verb|momentuHMMData| object
  \item{\verb|nbStates|} Number of latent states $(N)$
  \item{\verb|dist|} A list indicating the probability distributions for the data streams
  \item{\verb|estAngleMean|} A list indicating whether or not to estimate the mean for angular data streams (e.g.\ turning angle)
  \item{\verb|formula|} Model formula for state transition probabilities $(\gamma_{ij}^{(t)})$
  \item{\verb|stationary|} Logical indicating whether or not the initial distribution $({\boldsymbol \delta}^{(0)})$ is considered equal to the stationary distribution
  \item{\verb|Par0|} A list of starting values for the probability distribution parameters of each data stream
\end{itemize}
These seven arguments are all that are needed in order to fit the HMMs currently supported in \verb|moveHMM| \citep{MichelotEtAl2016}, but the similarities between the packages largely end here. Many of the arguments in \verb|fitHMM| and other \verb|momentuHMM| functions are lists, with each element of the list corresponding to a data stream. %The list names provided in \verb|dist|, \verb|Par0|, and \verb|estAngleMean| must therefore have a corresponding column in \verb|data| with the same name (e.g.\ `step' and `angle'). 
Additional data streams can be added to HMMs by simply adding additional elements to these list arguments (see section \ref{sec:nfs}).  

The \verb|formula| argument can include many of the functions and operators commonly used to construct terms in R model formulas (e.g.\ \verb|a*b|, \verb|a:b|, \verb|cos(a)|). It can also be used to specify transition probability matrix models that incorporate cyclical patterns using the \verb|cosinor| function (see section \ref{sec:elephant}), splines for explaining other more complicated patterns (e.g.\ \verb|bs| and \verb|ns| functions in the R base package \verb|splines|), and factor variables (e.g.\ group- or individual-level effects).  By default the \verb|formula| argument applies to all entries in the transition probability matrix $({\mathbf \Gamma}^{(t)})$, but special functions allow for state- and parameter-specific formulas to be specified% (see section \ref{sec:greySeal})
. Specific state transition probabilities can also be fixed to zero (or any other value), which can be useful for incorporating more behavioural realism by prohibiting or enforcing switching from one particular state to another (possibly as a function of spatio-temporal covariates). Similar arguments are available for %specifying covariate models and fixing parameters for the 
initial distribution parameters% $({\boldsymbol \delta}^{(0)})$
. 

The \verb|DM| argument of \verb|fitHMM| allows covariate models to be specified for the state-dependent probability distribution parameters of each data stream. %\verb|DM| is a list argument containing an element for each data stream, but each element itself is also a list specifying the design matrix formulas for each parameter. 
\verb|DM| formulas are just as flexible as the \verb|formula| argument. %and, in addition to common linear model formula functions and operators, can also include cyclical cosinor models, splines, factor variables, and state-specific probability distribution parameter formulas 
%(see sections \ref{sec:elephant}, \ref{sec:turtle}, and \ref{sec:greySeal}). However, specification of design matrices for probability distribution parameters using \verb|DM| is not limited to formulas. 
In lieu of formulas, \verb|DM| can also be specified as a ``pseudo-design'' matrix with rows and columns corresponding to the natural and working scale parameters, respectively (Table \ref{tab:pdfs}). %, and these enable parameter constraints that can be particularly useful for preventing state label switching when using multiple imputation methods (see section \ref{sec:mi}). %Probability distribution parameters can also be fixed to user-specified values.
Pseudo-design matrices allow working parameters to be shared among natural parameters for imposing relational constraints (e.g.\ $\mu_1 = \mu_2$, $\mu_1 \le \mu_2$). Probability distribution parameters can also be fixed to user-specified values.

Another noteworthy \verb|fitHMM| argument, \verb|circularAngleMean|, enables circular-circular regression models for the mean direction of angular distributions, such as the wrapped Cauchy and von Mises, instead of circular-linear models based on the tangent link function (Table \ref{tab:pdfs}). %When \verb|circularAngleMean| is specified for any given angular data stream (e.g.\ turning angle), then a special link function based on \cite{DuchesneEtAl2015} is used:
%\begin{equation}
%  {\boldsymbol \mu}=\text{atan2}(\sin({\mathbf X}_\mu){\boldsymbol \beta}_\mu,1+\cos({\mathbf X}_\mu){\boldsymbol \beta}_\mu),
%  \label{eq:circ}
%\end{equation}
%where ${\mathbf X}_\mu$ is a $T \times k$ matrix composed of the turning angles between $k$ angular covariates and the bearing of movement during the %previous time step, i.e., each element 
%\begin{equation}
%x_{t,i}=\text{atan2}(\sin(r_{t,i}-b_{t-1}),\cos(r_{t,i}-b_{t-1})) 
%  \label{eq:angleCov}
%\end{equation}
%for angular covariate $r_{t,i}$ and $i=1,\ldots,k$%(note that \verb|prepData| calculates ${\mathbf X}_\mu$ based on the \verb|angleCovs| or \verb|centers| arguments so users need not bother)
%. Because this link function is designed for turning angles, a turning angle of 0 is provided as the reference angle% (hence the ``$1+$'' preceeding the cosine term in Eq.\ \ref{eq:circ}).  Thus as a trade-off between biased and correlated movements, the working parameters $({\boldsymbol \beta}_\mu)$ for the expected turning angle at time $t$ weight the attractive (or repulsive) strengths of the angular covariates relative to directional persistence.  When all ${\boldsymbol \beta}_\mu=0$, the model reduces to a correlated random walk, but an increasingly biased random walk results as ${\boldsymbol \beta}_\mu$ gets larger (in absolute value). Many interesting hypotheses can be addressed using circular-circular regression, including the effects of wind, sea surface currents, and centres of attraction (or repulsion) on animal movement direction (see examples in section \ref{sec:turtle} and \ref{sec:greySeal}).
As a trade-off between biased and correlated movements, circular-circular regression models can weight the attractive (or repulsive) strengths of angular covariates relative to short-term directional persistence using a special link function based on \cite{RivestEtAl2016}. Many interesting hypotheses can be addressed using circular-circular regression, including the effects of wind, sea surface currents, and centres of attraction on animal movement direction (see sections \ref{sec:turtle} and \ref{sec:greySeal}). 

\subsection{Multiple imputation}
\label{sec:mi}
When location data are temporally-irregular or subject to measurement error, %then they are not suitable for standard maximum-likelihood HMM analyses based on the forward algorithm. In this case, 
\verb|momentuHMM| can be used to perform the 2-stage multiple imputation approach of \cite{McClintock2017}. %The basic concept is to first employ a single-state (i.e.\ $N=1$) movement model that is relatively easy to fit but can accommodate location measurement error and temporally-irregular or missing observations \citep[e.g.][]{JohnsonEtAl2008}. The second stage involves repeatedly fitting the desired HMM to $m$ temporally-regular realizations of the position process drawn from the model output of the first stage.  Data streams or covariates that are dependent on location (e.g.\ step length, turning angle, habitat type, snow depth, sea surface temperature) will of course vary among the $m$ realizations of the position process, and the pooled inferences across the HMM analyses therefore reflect location uncertainty.  
There are three primary functions for performing multiple imputation HMM analyses: \verb|crawlWrap|, \verb|MIfitHMM|, and \verb|MIpool|%, and all rely on parallel processing to speed up computations
. Based on the R package \verb|crawl| \citep{Johnson2017}, \verb|crawlWrap| is a wrapper function for fitting the continuous-time correlated random walk (CTCRW) model of \cite{JohnsonEtAl2008} and then predicting temporally-regular tracks of the user's choosing (e.g.\ 15 min, hourly, daily). %\verb|crawlWrap| returns a \verb|crwData| object that can be used to draw $m$ realization of the position process within the \verb|MIfitHMM| function.  
\verb|MIfitHMM| is essentially a wrapper function for \verb|fitHMM| that repeatedly fits the same user-specified HMM to $m$ imputed data sets using maximum likelihood% and stores the output from each of the $m$ model fits. If a \verb|crwData| object is provided, then \verb|MIfitHMM| will first draw $m$ imputations based on the \verb|crwData| output and then fit the specified HMM to each imputed data set. If users wish to use a movement model other than the CTCRW to account for measurement error and temporal irregularity \citep[e.g.][]{CalabreseEtAl2016,GurarieEtAl2017}, or if other observation error processes (e.g.\ missing data) are to be accounted for in the imputation step, \verb|MIfitHMM| can also be used for analysis of a list of $m$ \verb|momentuHMMData| objects that were imputed by the user
. Based on the $m$ model fits, the \verb|MIpool| function calculates pooled estimates, standard errors, and confidence intervals %for the working scale parameters, natural scale parameters, state sequences, state probabilities, and activity budgets (i.e.\ the proportion of the $T$ times step assigned to each state) 
using standard multiple imputation formulae. % \verb|MIpool| can be called separately or within \verb|MIfitHMM| (using the \verb|poolEstimates| argument), and the function returns a \verb|miSum| object containing the pooled output across all imputatons. 
Data streams or covariates that are dependent on location (e.g.\ step length, turning angle, habitat type, snow depth, sea surface temperature) will of course vary among the $m$ realizations of the position process, and the pooled inferences across the fitted HMMs therefore reflect location uncertainty. %See sections \ref{sec:nfs}, \ref{sec:turtle}, and \ref{sec:greySeal} for example HMM analyses using multiple imputation to account for location measurement error and temporal irregularity.

\subsection{Model visualization and diagnostics}
The generic \verb|plot| function for \verb|momentuHMM| model objects plots the data stream histograms along with their corresponding estimated probability distributions, the estimated natural parameters and state transition probabilities as a function of any covariates included in the model, and the tracks of all individuals (colour-coded by the most likely state sequence). %By default, the probability distributions are plotted based on the means of any covariate values, but user-specified covariate values for the plots can be provided using the \verb|covs| argument.  
Confidence intervals for the natural parameters and state transition probabilities can also be plotted. %Confidence intervals are calculated from the working parameter estimates based on the delta method and finite-difference approximations of the first derivative for the transformation using the \verb|numDeriv::grad| function \citep{GilbertVaradhan2016}.  
For multiple imputation analyses, estimated 95\% location error ellipses can be included in plots of individual tracks. %The functions \verb|plotSat|, \verb|plotSpatialCov|, and \verb|plotStates| (Table \ref{tab:functions}) provide further methods for visualizing model results. 
Diagnostic tools include the calculation and plotting of pseudo-residuals \citep{ZucchiniEtAl2016} using the \verb|pseudoRes| and \verb|plotPR| functions, respectively. %For discrete distributions (e.g.\ Bernoulli, Poisson), a continuity adjustment is used for calculating pseudo-residuals. 
Akaike's Information Criterion can be calculated for one or more models using the \verb|AIC.momentuHMM| function.

\subsection{Simulation}
The function \verb|simData| can be used to simulate multivariate HMMs from scratch or from a fitted model. The arguments are very similar to those used for data preparation and model specification, but they include additonal arguments for simulating location data subject to temporal irregularity and measurement error. Among other features, activity centres and rasters of spatio-temporal covariates can be utilized (see section \ref{sec:greySeal}). \verb|simData| can therefore simulate more ecologically-realistic tracks (potentially subject to observation error) 
for study design, power analyses, and assessing model performance or goodness-of-fit \citep{MoralesEtAl2004}. %Goodness-of-fit can also be investigated by drawing simulated data sets from a fitted model and comparing them to observed properties of the data \citep{MoralesEtAl2004}.
  
\section{Examples}
\label{sec:example}
We now demonstrate some of the capabilities of \verb|momentuHMM| using real telemetry data. Intended for demonstration purposes only, we do not claim these represent improvements relative to previous or alternative analyses for these data sets. While space limits us to only a few key elements here, complete details and code for reproducing all examples described herein are provided in the ``vignettes'' source directory.

\subsection{African elephant}
\label{sec:elephant}
\begin{figure}[htbp]
  \includegraphics[width=0.49\textwidth]{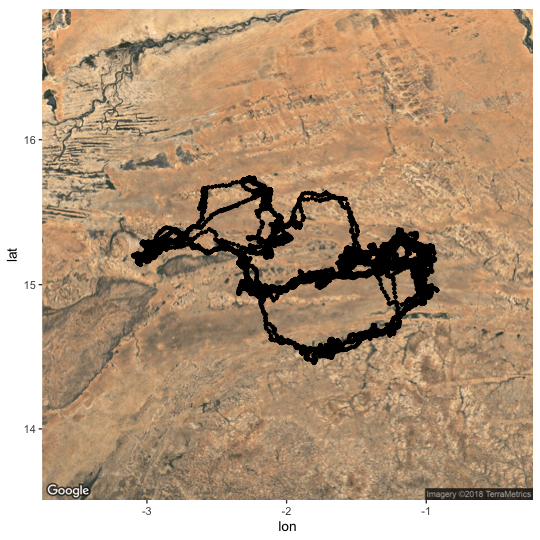}
  \includegraphics[width=0.49\textwidth]{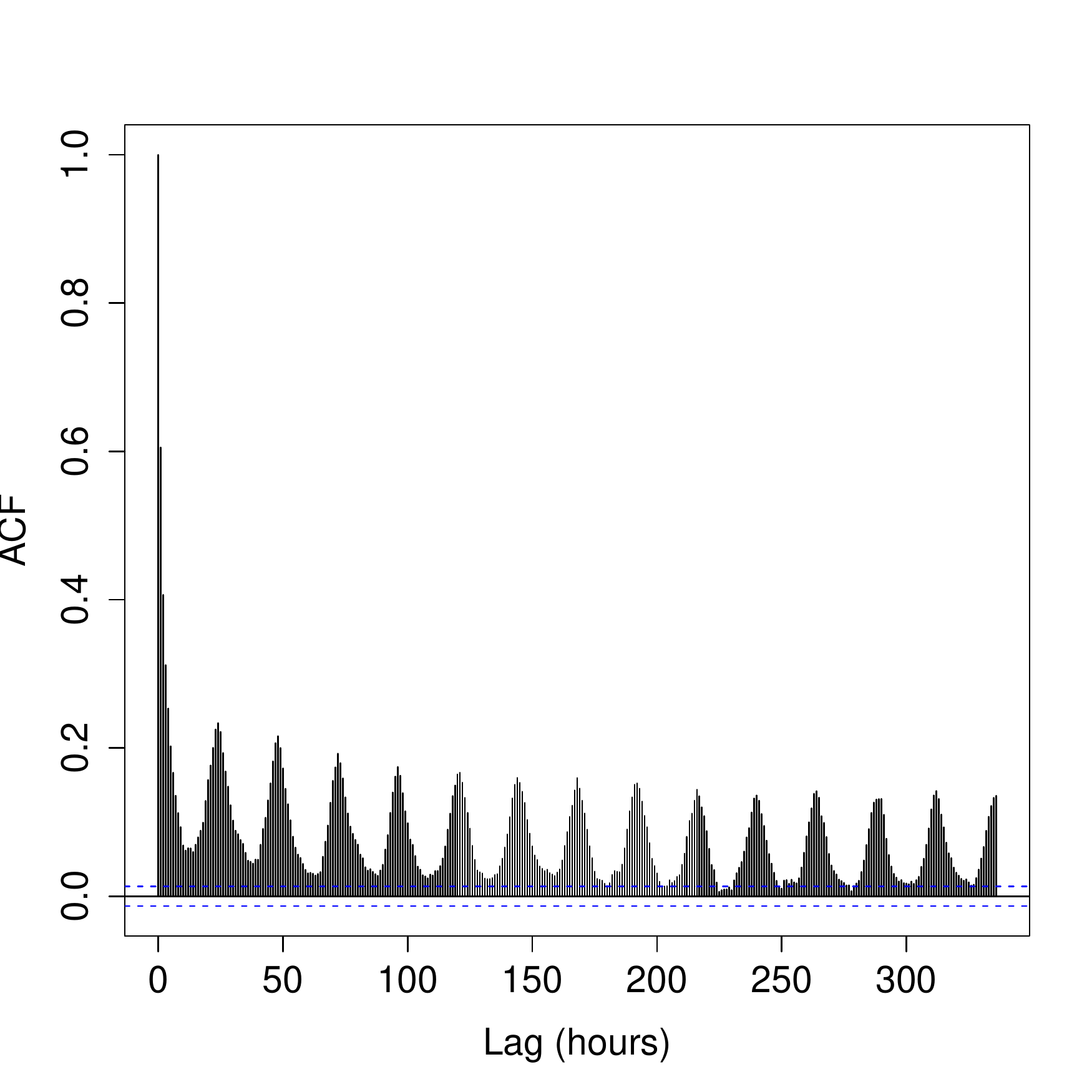}
  \includegraphics[width=0.49\textwidth]{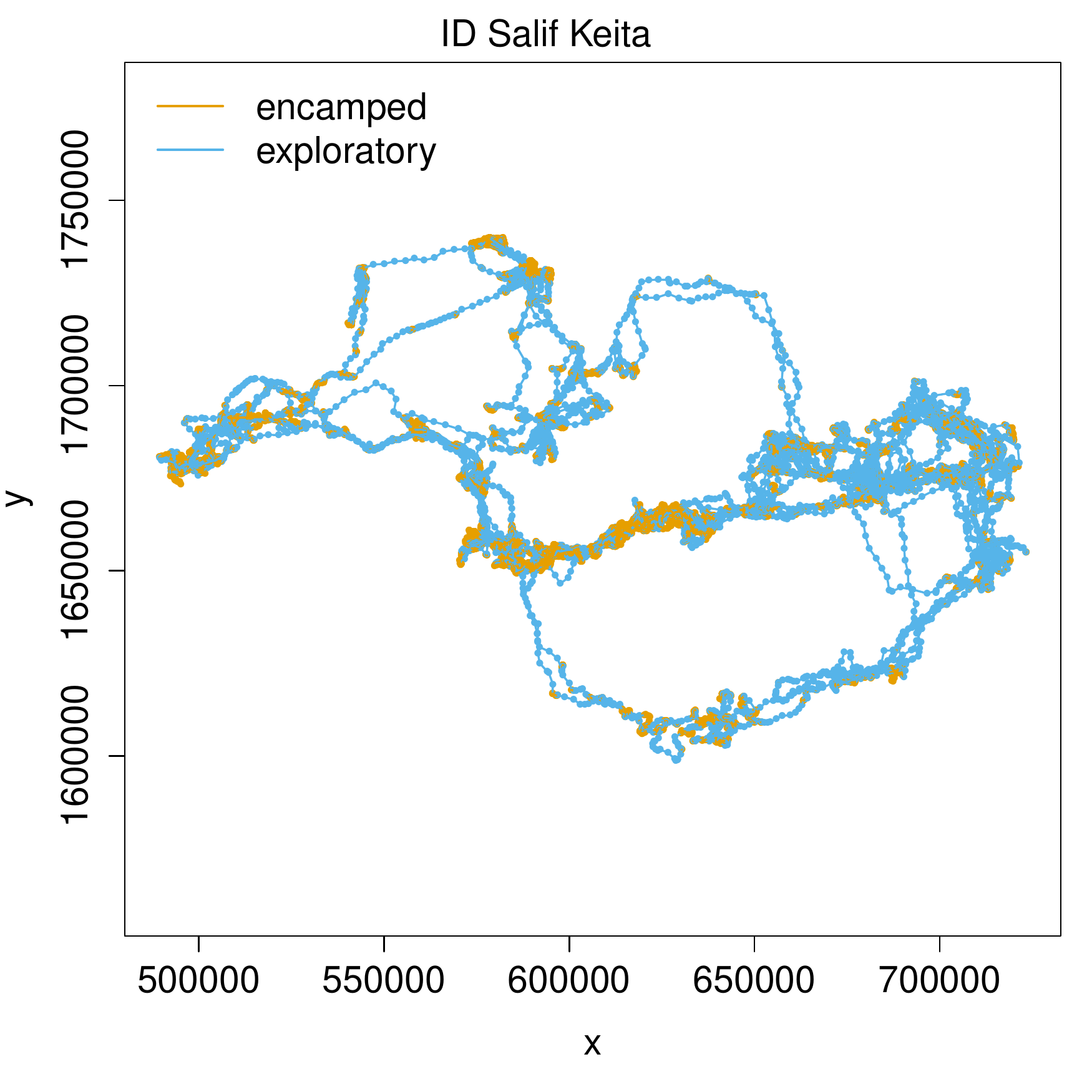}
  \includegraphics[width=0.49\textwidth]{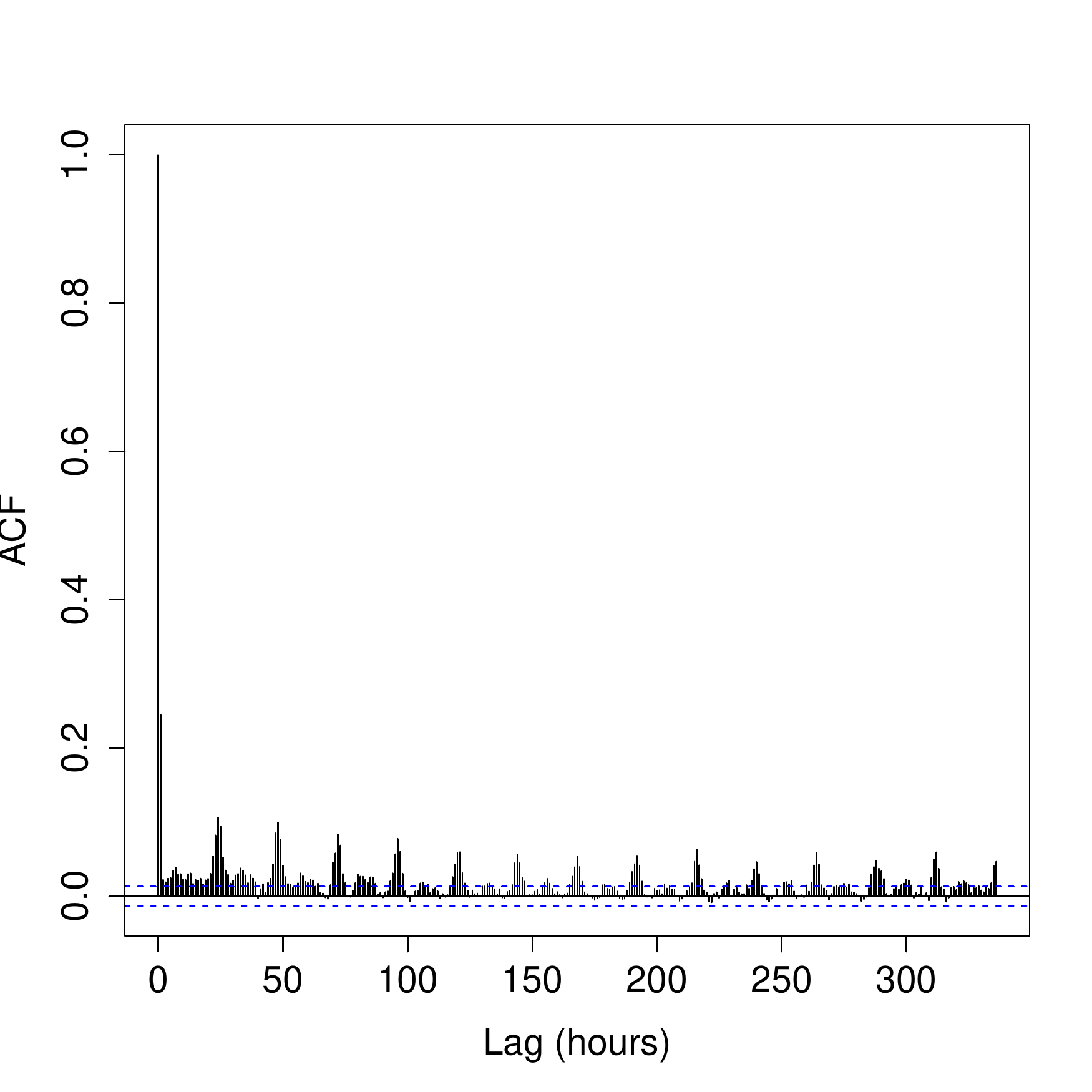}
  \caption{Selected plots for the 2-state (``encamped'' and ``exploratory'') African elephant example% generated using the generic 'plot' and 'acf' functions
. About 74\% and 26\% of the 1 h time steps were attributed to the ``encamped'' and ``exploratory'' states, respectively. Top-left panel is a plot of the elephant track produced using the `plotSat' function, top-right panel presents the autocorrelation function (ACF) plot of the step length data, bottom-left panel is a plot of the Viterbi-decoded track (orange = ``encamped'', blue = ``exploratory'') generated from the fitted model using the generic `plot' function, and bottom-right panel presents the step length pseudo-residual ACF plot using the `plotPR' function.}
  \label{fig:elephantResults1}
\end{figure}

As our first example, we use an African elephant ({\it Loxodonta africana}) bull track described in \cite{WallEtAl2014}% and publicly available from the \url{movebank.org} data repository. The data set contains two tracks; for this analysis, we only consider the first one
. In addition to hourly locations, the tag collected external temperature data. %Location measurement error is negligible for these terrestrial GPS data, although about 1\% of the hourly observations collected between 22 March 2008 and 30 September 2010 are missing. Instead of simply ignoring these missing data, we employed \verb|crawlWrap| to predict the missing hourly locations prior to fitting a 2-state HMM assuming a wrapped Cauchy distribution for turning angle and a gamma distribution for step length. 
Using \verb|fitHMM|, we fitted a 2-state HMM assuming a wrapped Cauchy distribution for turning angle and a gamma distribution for step length. Autocorrelation function estimates suggest there are 24 h cycles in the step length data (Fig.\ \ref{fig:elephantResults1}), and this presents an opportunity to demonstrate the cosinor model for incorporating cyclical behaviour in model parameters. We therefore specified temperature effects on the turning angle concentration parameters and cycling temperature effects (with a 24 h periodicity) on the step length and state transition probability parameters using the \verb|DM| and \verb|formula| arguments. 

\begin{figure}[htbp]
  \centering
  \includegraphics[width=0.35\textwidth]{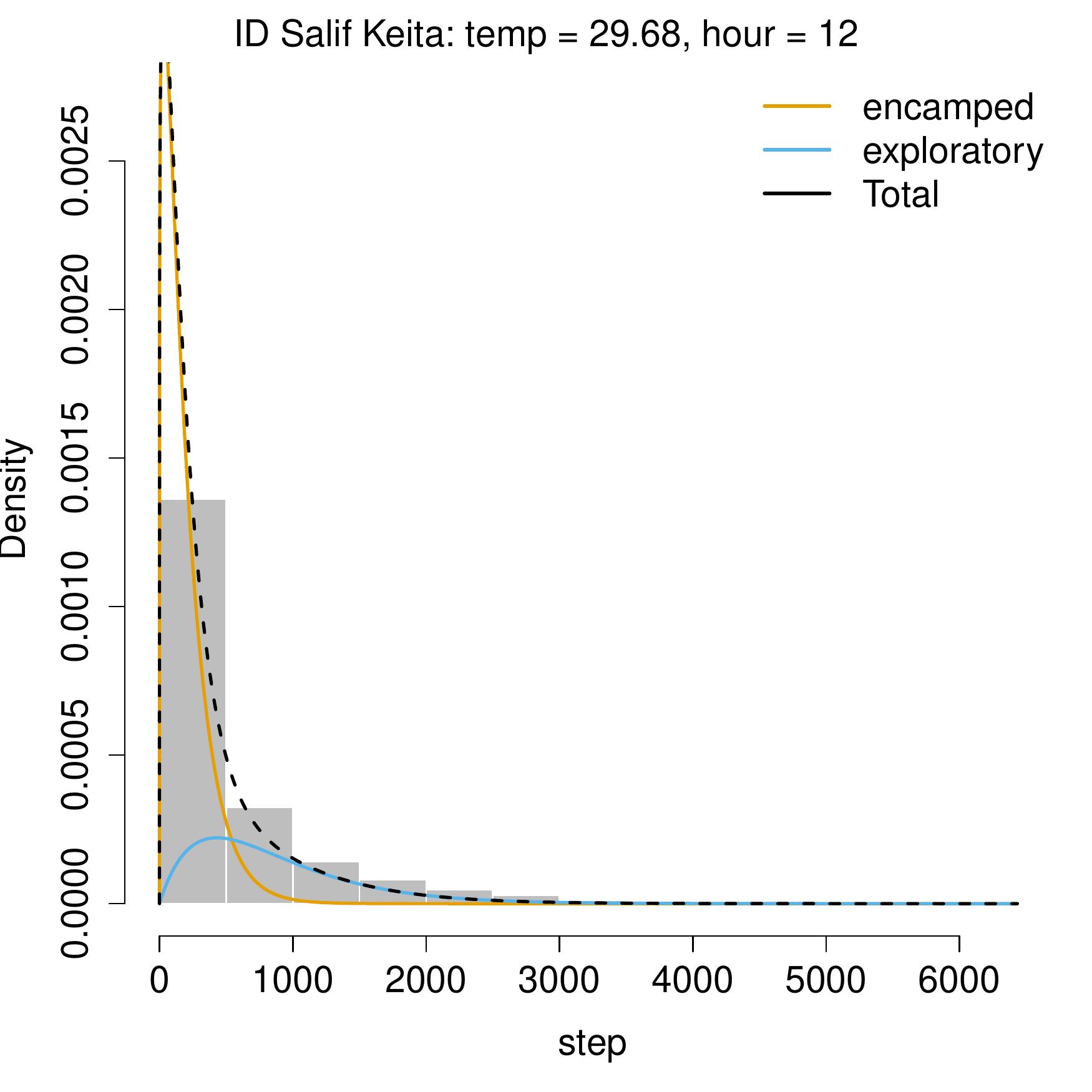}
  \includegraphics[width=0.35\textwidth]{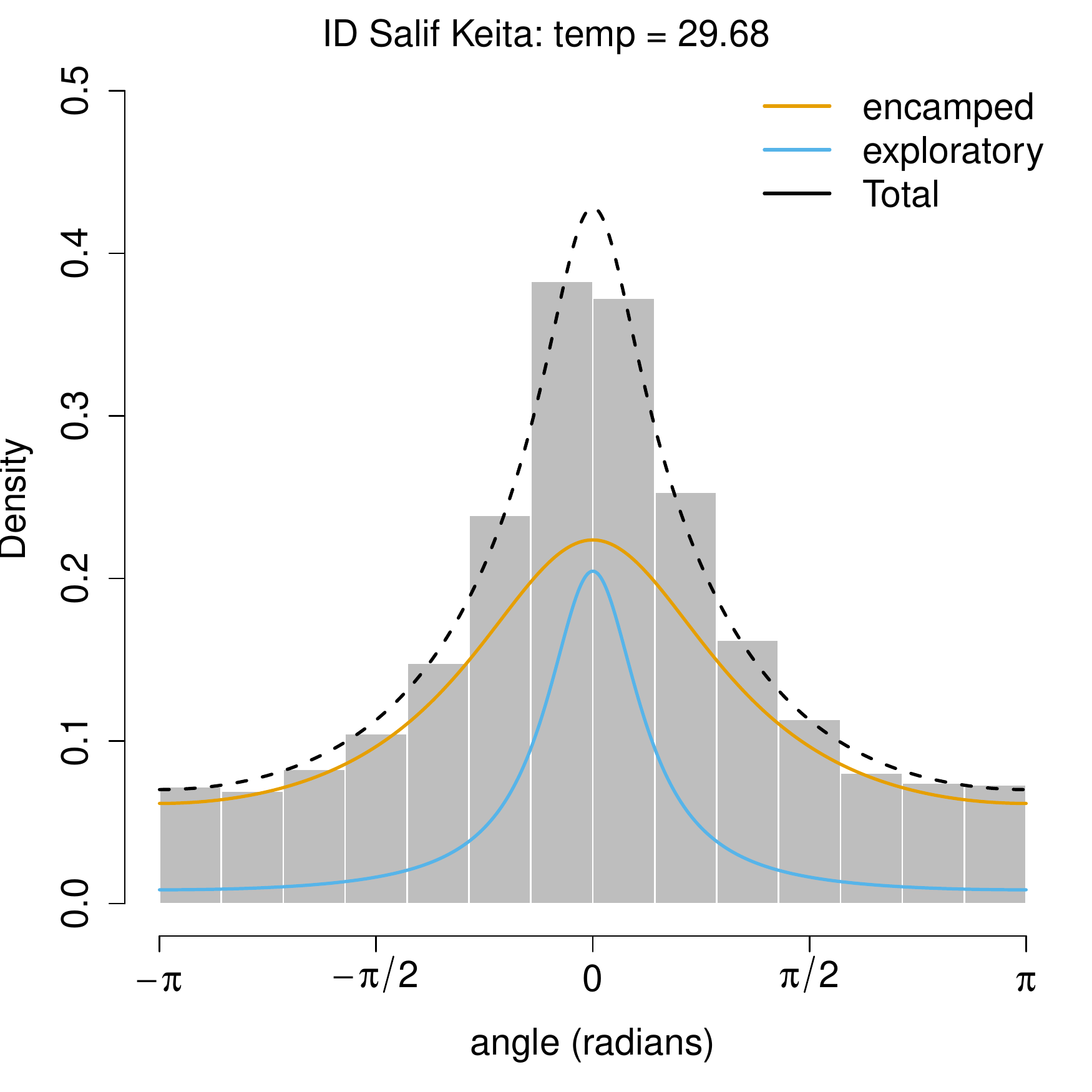} \\
  \includegraphics[width=0.35\textwidth]{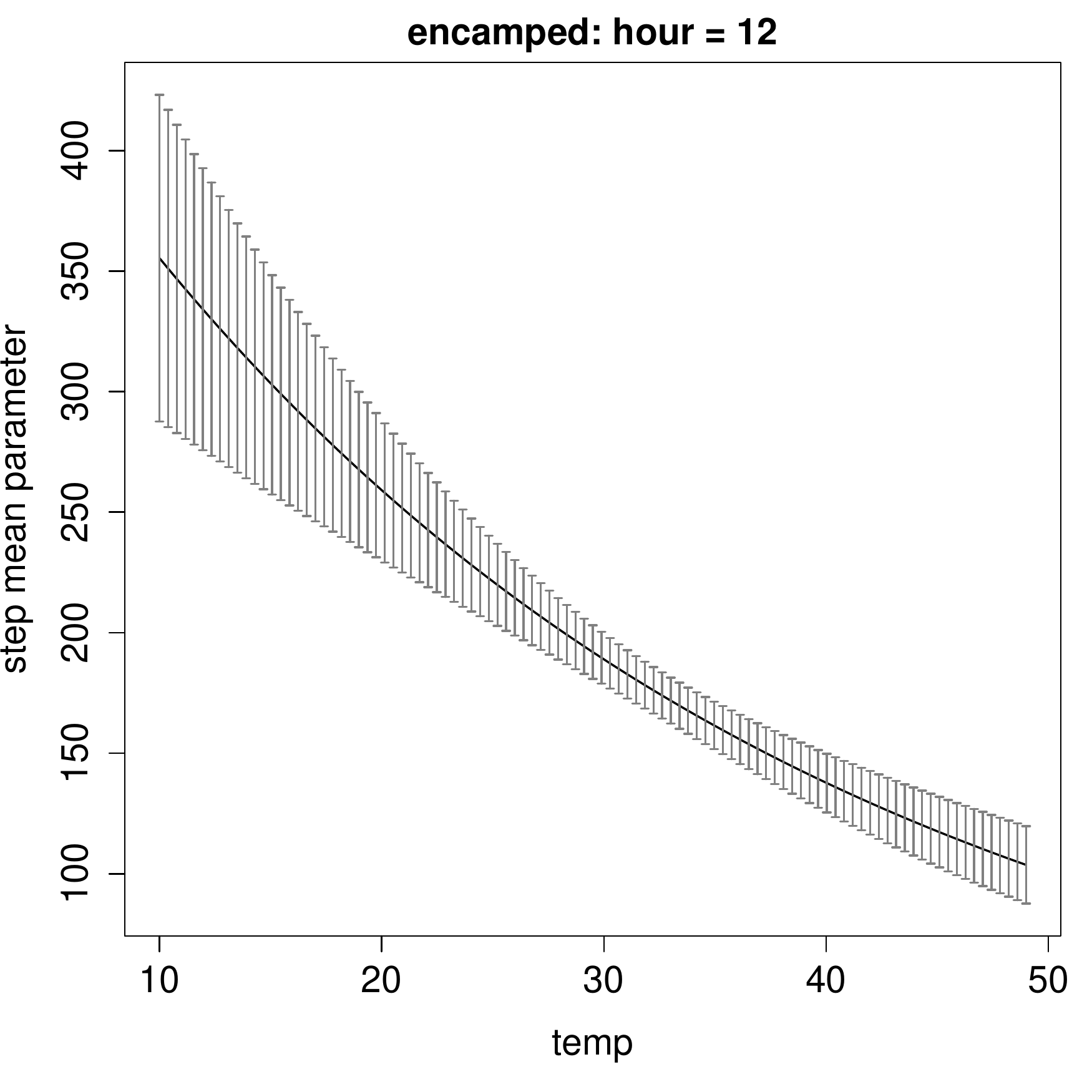} 
  \includegraphics[width=0.35\textwidth]{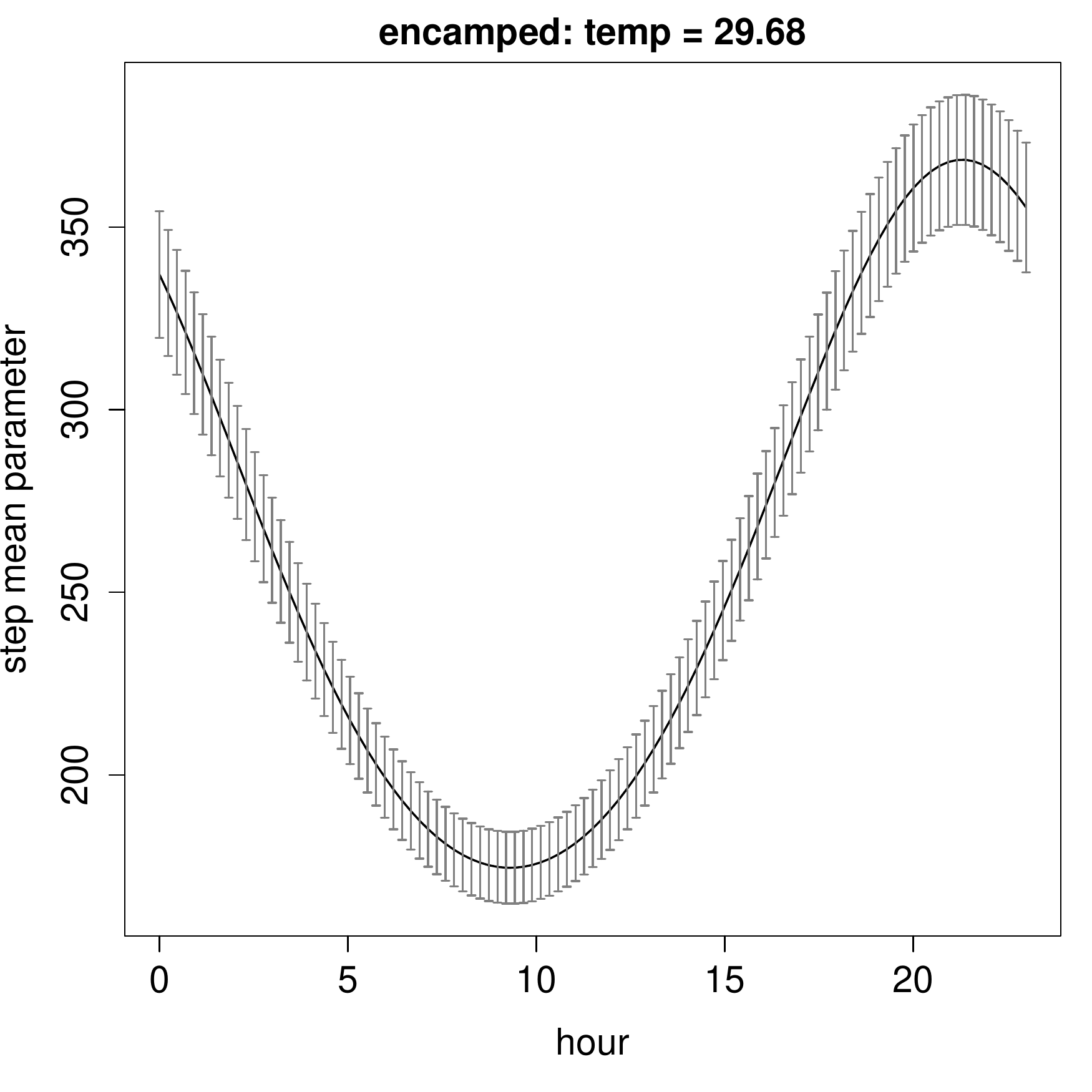} \\
  \includegraphics[width=0.35\textwidth]{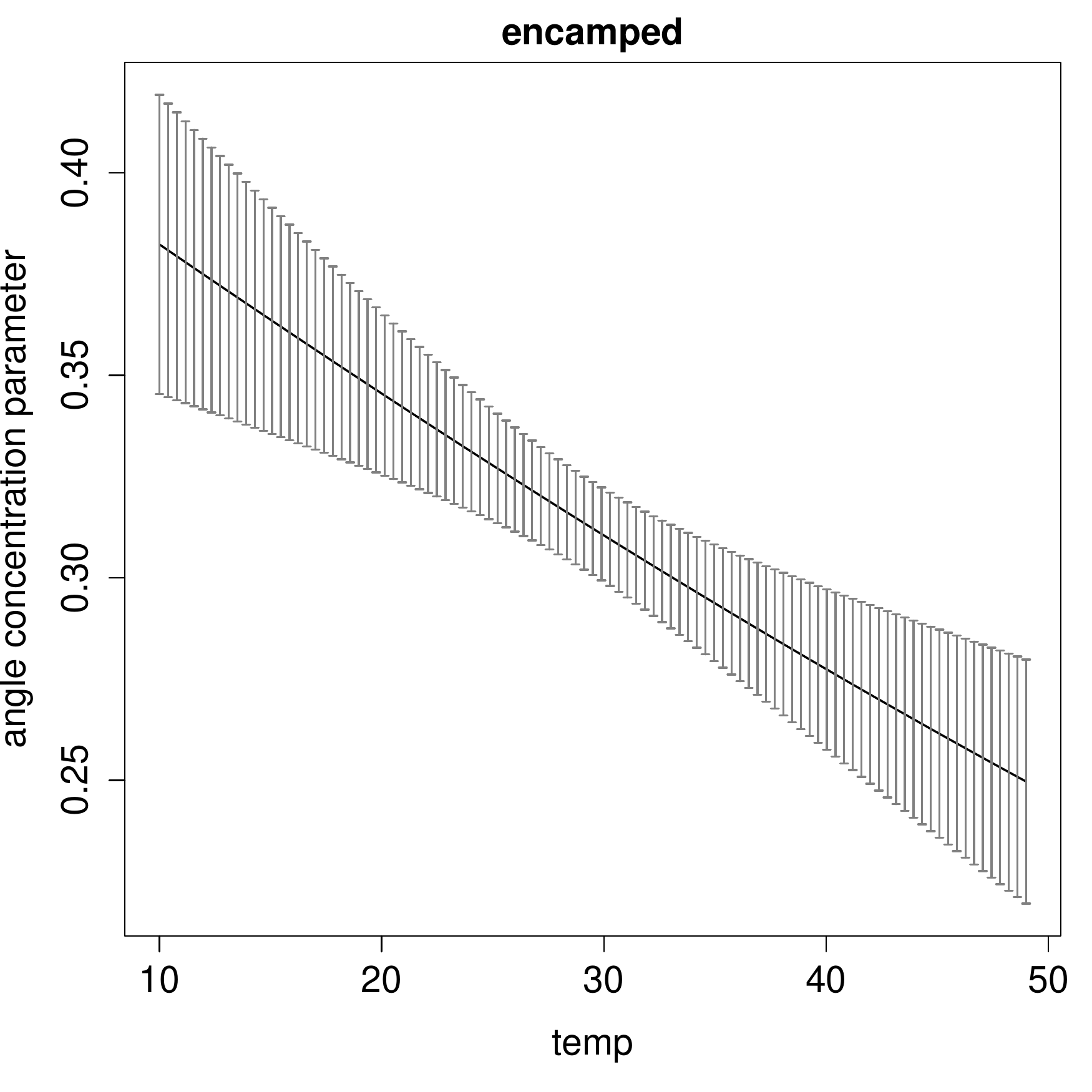}
  \includegraphics[width=0.35\textwidth]{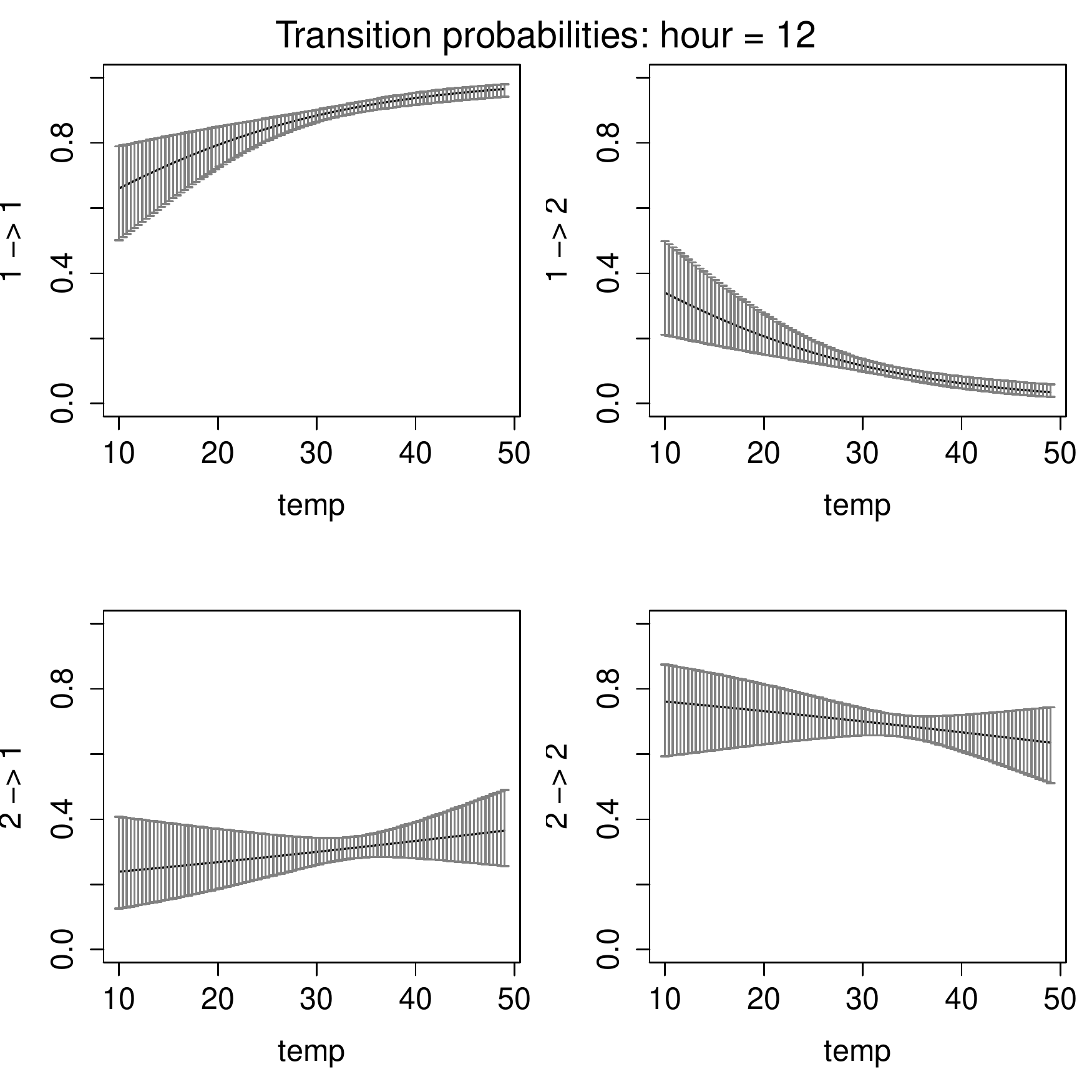} \\
  \caption{Estimated probability distribution and parameter plots for the 2-state (``encamped'' and ``exploratory'') African elephant example generated using the generic 'plot' function. Top panels present histograms of the step length (top-left) and turning angle (top-right) data along with the estimated state-dependent probability distributions based on the mean temperature (temp = 29.7 degrees celsius) at 12:00 GMT (hour = 12). Middle panels present estimates (and 95\% confidence intervals) for the step length mean parameter of the ``encamped'' state as a function of temperature and hour of day.  Bottom-left panel presents estimates for the turning angle concentration parameter of the ``encamped'' state as a function of temperature.  Bottom-right panel presents estimated state transition probabilities (1 = ``encamped'', 2 = ``exploratory'') as a function of temperature at 12:00 GMT.}
  \label{fig:elephantResults2}
\end{figure}

The model identifed a state of slow undirected movement (``encamped''), and a state of faster and more directed movement (``exploratory'') (Fig.\ \ref{fig:elephantResults1}). %About 74\% of the steps were attributed to the ``encamped'' state, and 26\% were attributed to the ``exploratory'' state. 
Interestingly, this model suggests step lengths and directional persistence for the ``encamped'' state decreased as temperature increased, step lengths for both states tended to decrease in the late evening and early morning, and transition probabilities from the ``encamped'' to ``exploratory'' state decreased as temperature increased (Fig.\ \ref{fig:elephantResults2}). This model was overwhelmingly supported by AIC when compared to alternative models with fewer covariates, but model fit can be further assessed using pseudo-residuals.  In this case, the pseudo-residuals indicate the model explained much of the periodicity in step length, although there does still appear to be some room for improvement (Fig.\ \ref{fig:elephantResults1}).

\subsection{Northern fur seal}
\label{sec:nfs}
We use the northern fur seal ({\it Callorhinus ursinus}) example from \cite{McClintockEtAl2014b} to demonstrate the use of additional data streams for distinguishing behaviours with similar horizontal trajectories (e.g.\ ``resting'' and ``foraging'') in a multivariate HMM. The data consist of 241 temporally-irregular Fastloc GPS locations obtained during a foraging trip of a nursing female near the Pribilof Islands of Alaska, USA. The tag included time-depth recording capabilities, and the dive activity data were summarized as the number of foraging dives over $T=228$ temporally-regular 1 h time steps. To fit the $N=3$ state (1=``resting'', 2=``foraging'', 3=``transit'') model of \cite{McClintockEtAl2014b}, we first used \verb|crawlWrap| to predict temporally-regular locations at 1 h time steps assuming a bivariate normal location measurement error model. %and merged the results with the foraging dive data using the \verb|crawlMerge| function. 
We then used multiple imputation to account for location uncertainty by repeatedly fitting the 3-state HMM to $m=100$ realizations of the position process using \verb|MIfitHMM|. We specified a gamma distribution for step length, wrapped Cauchy distribution for turning angle, and Poisson distribution for the number of foraging dives. %We used several arguments in \verb|MIfitHMM| to avoid label switching among the 100 imputed data model fits and enforce similar state-dependent probability distribution constraints as \cite{McClintockEtAl2014b}, e.g.\ constraining the Poisson rate parameters such that the ``foraging'' state tends to have more foraging dives than the ``transit'' state ($\lambda_2 > \lambda_3$). %To prohibit foraging dives for the ``resting'' state, we used the \verb|fixPar| argument to effectively fix the Poisson rate parameter to zero on the natural scale (i.e.\ $\lambda_1 \approx 0$).

\begin{figure}[htbp]
  \includegraphics[width=0.49\textwidth,page=1]{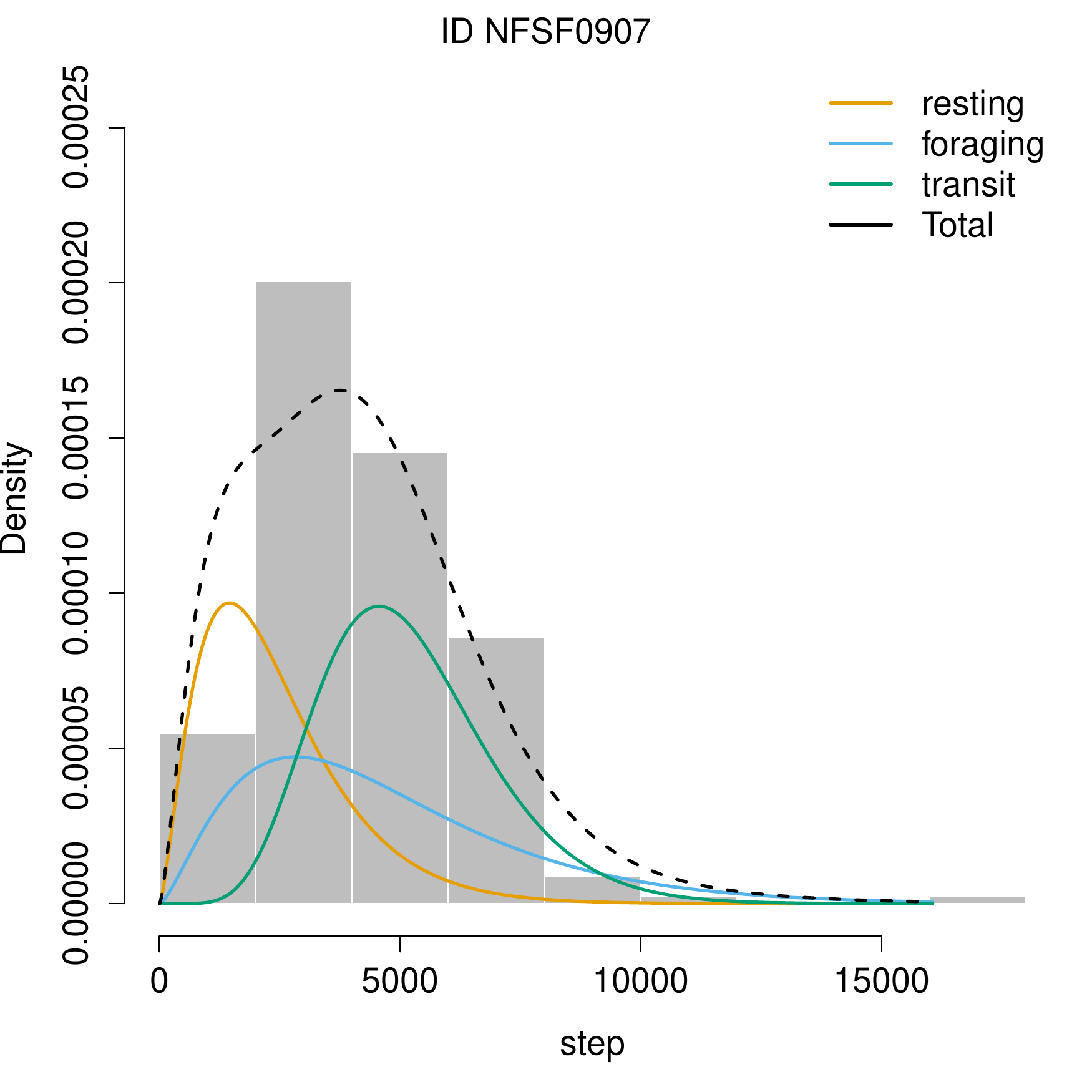}
  \includegraphics[width=0.49\textwidth,page=2]{plot_nfsResults}
  \includegraphics[width=0.49\textwidth,page=3]{plot_nfsResults}
  \includegraphics[width=0.49\textwidth,page=4]{plot_nfsResults}
  \caption{Plots of the northern fur seal example results generated using the generic `plot' function. The estimated probability distributions for step length (top-left panel), turning angle (top-right panel), and number of foraging dives (bottom-left panel) for the 3-state (``resting'', ``foraging'', and ``transit'') model are plotted along with histograms of these data streams. The temporally-regular predicted locations, 95\% ellipsoidal confidence bands, and estimated states are plotted in the bottom-right panel. All estimates are pooled across multiple imputations of the position process and thus reflect uncertainty attributable to location measurement error and temporally-irregular observations.}
  \label{fig:nfsResults}
\end{figure}

Our results were very similar to those of the discrete-time Bayesian model of \cite{McClintockEtAl2014b}, with periods of ``foraging'' often followed by ``resting'' at sea (Fig.\ \ref{fig:nfsResults}).  Activity budgets based on the estimated state sequences for each imputation were 0.31 (95\% CI: 0.23$-$0.4) for ``resting'', 0.28 (95\% CI: 0.23$-$0.34) for ``foraging'', and 0.41 (95\% CI: 0.31$-$0.51) for ``transit''.

\subsection{Loggerhead turtle}
\label{sec:turtle}
Using hitherto unpublished loggerhead turtle ({\it Caretta caretta}) data for a captive-raised juvenile released on the coast of North Carolina, USA, we now demonstrate how movement direction and step length can be easily modelled as a function of angular covariates in \verb|momentuHMM|. The data consist of %165 
temporally-irregular Argos locations and rasters of daily ocean surface currents collected between 20 November and 19 December 2012. Assuming a gamma distribution for step length %$(l_t)$ 
and a wrapped Cauchy distribution for turning angle%$(\phi_t)$
, we modelled the mean step length parameter %$(\mu^l_t)$ 
as a function of ocean surface current speed $(w_t)$ and direction $(r_t)$ relative to the bearing of movement $(b_t)$ at time $t$%:
%\begin{equation}
%  \mu^l_t=\exp(\beta^l_0+\beta^l_1 w_t \cos(b_t-r_t)),
%  \label{eq:turtleMeanStep}
%\end{equation}
. We also modelled the turning angle mean parameter %$(\mu^\phi_t)$ 
as a trade-off between short-term directional persistence and bias in the direction of ocean surface currents using the circular-circular regression link function.
%\begin{equation}
%  \mu^\phi_t=\text{atan2}(\sin(d_t) \beta^\phi,1+\cos(d_t)\beta^\phi),
%    \label{eq:turtleMeanAngle}
%\end{equation}
%where $d_t=\text{atan2}(\sin(r_t-b_{t-1}),\cos(r_t-b_{t-1}))$.

\begin{figure}[htbp]
  \centering
  \includegraphics[width=0.32\textwidth]{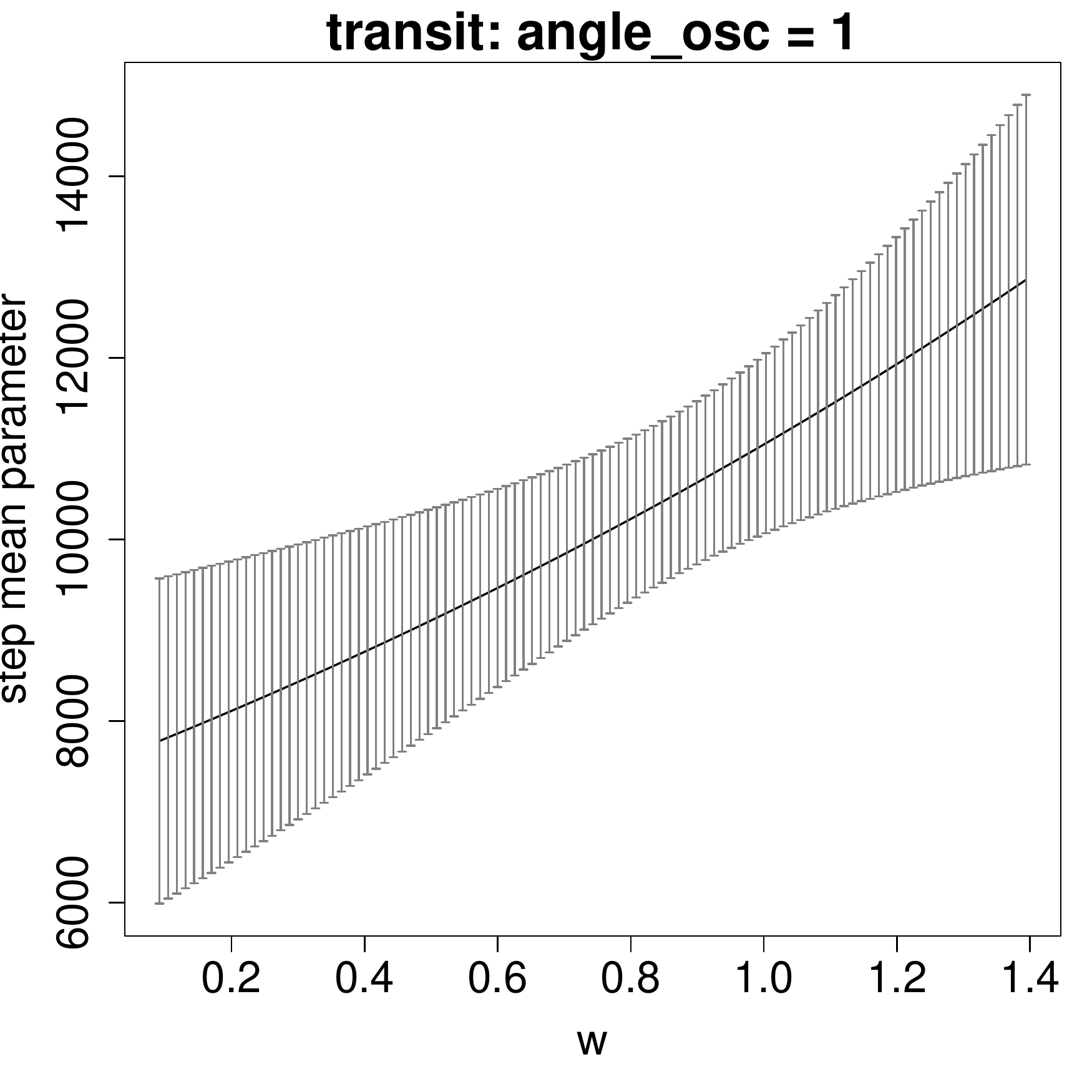}
  \includegraphics[width=0.32\textwidth]{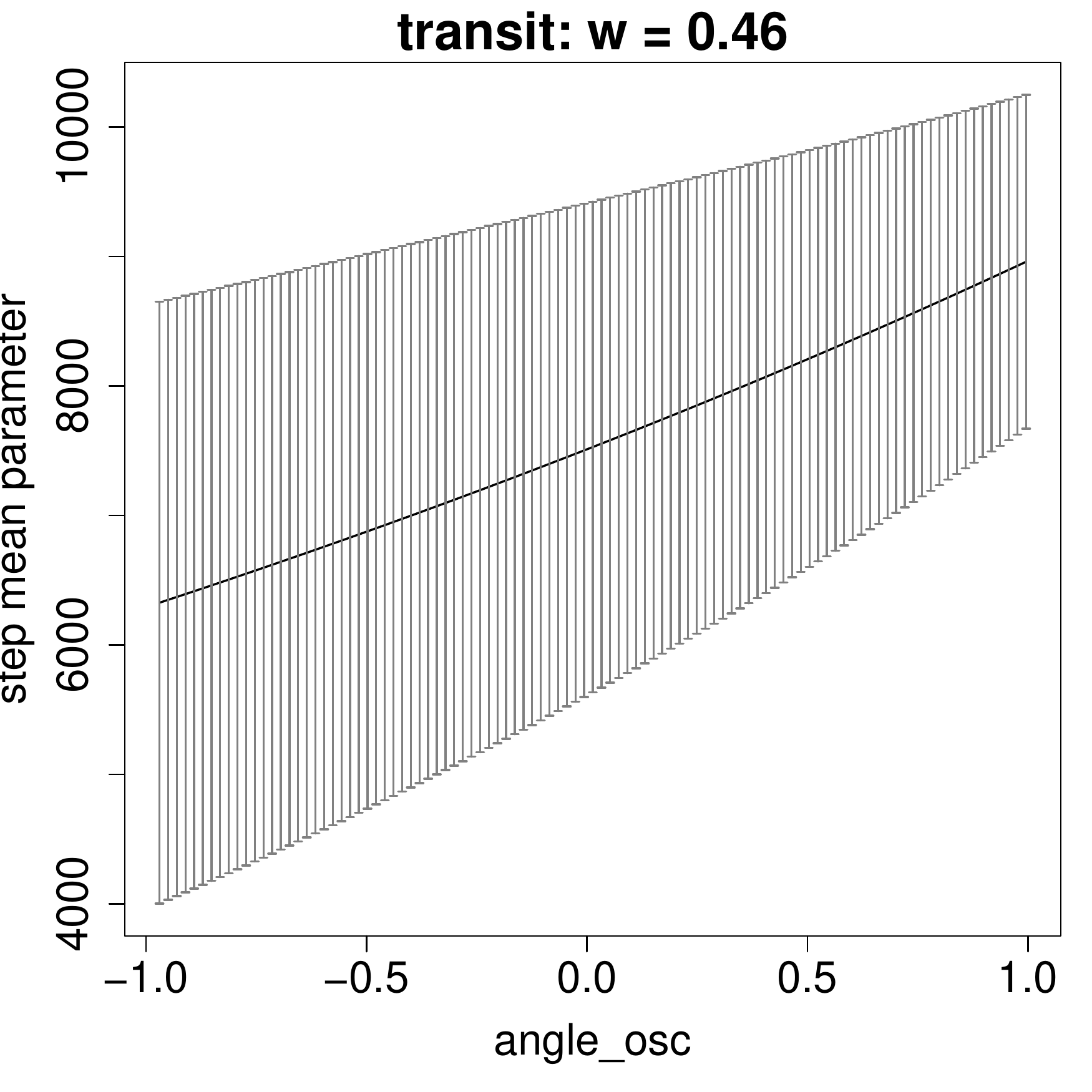}
  \includegraphics[width=0.32\textwidth]{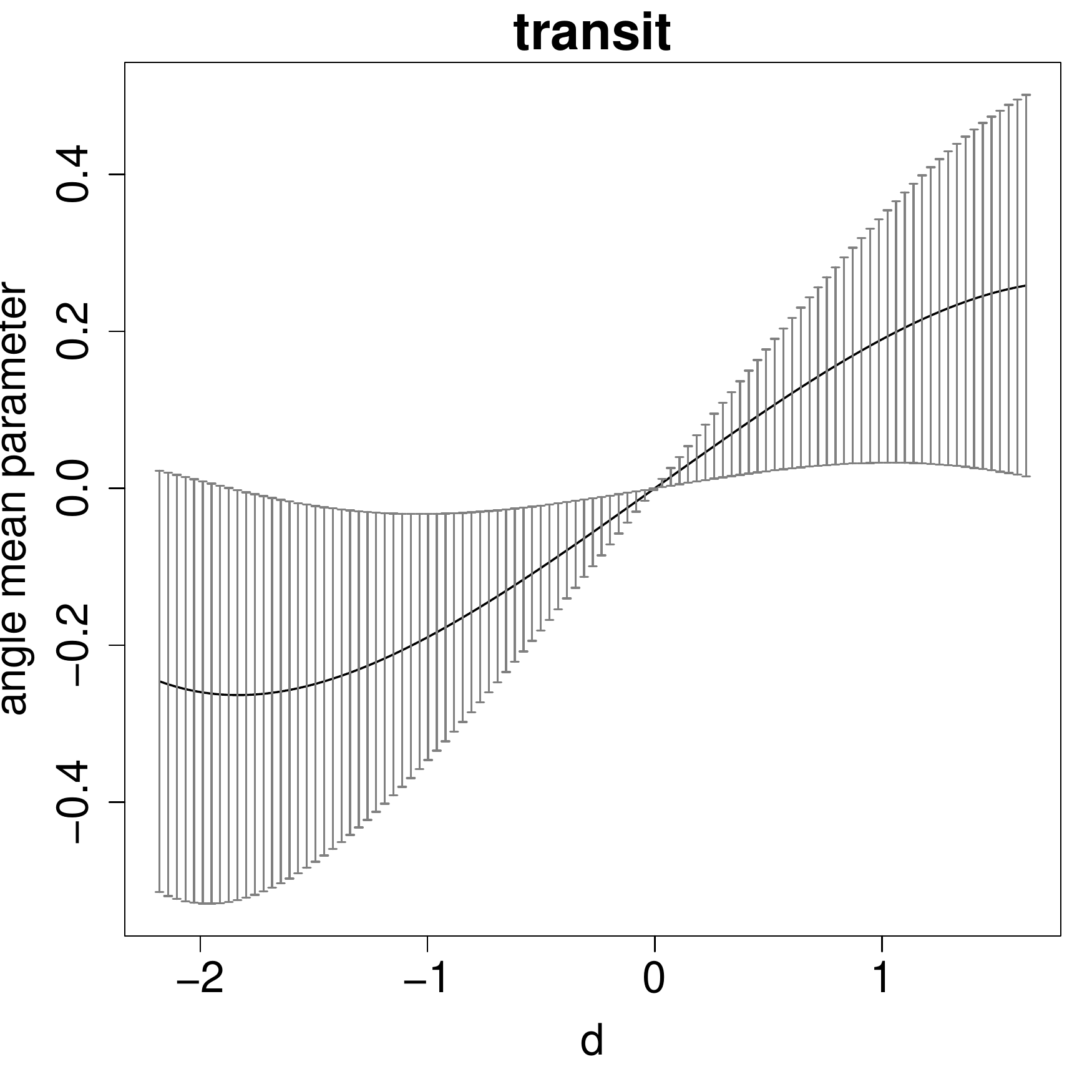}
  %\begin{adjustbox}{trim=0cm 0.25cm 0cm 1.5cm}
    \includegraphics[width=\textwidth]{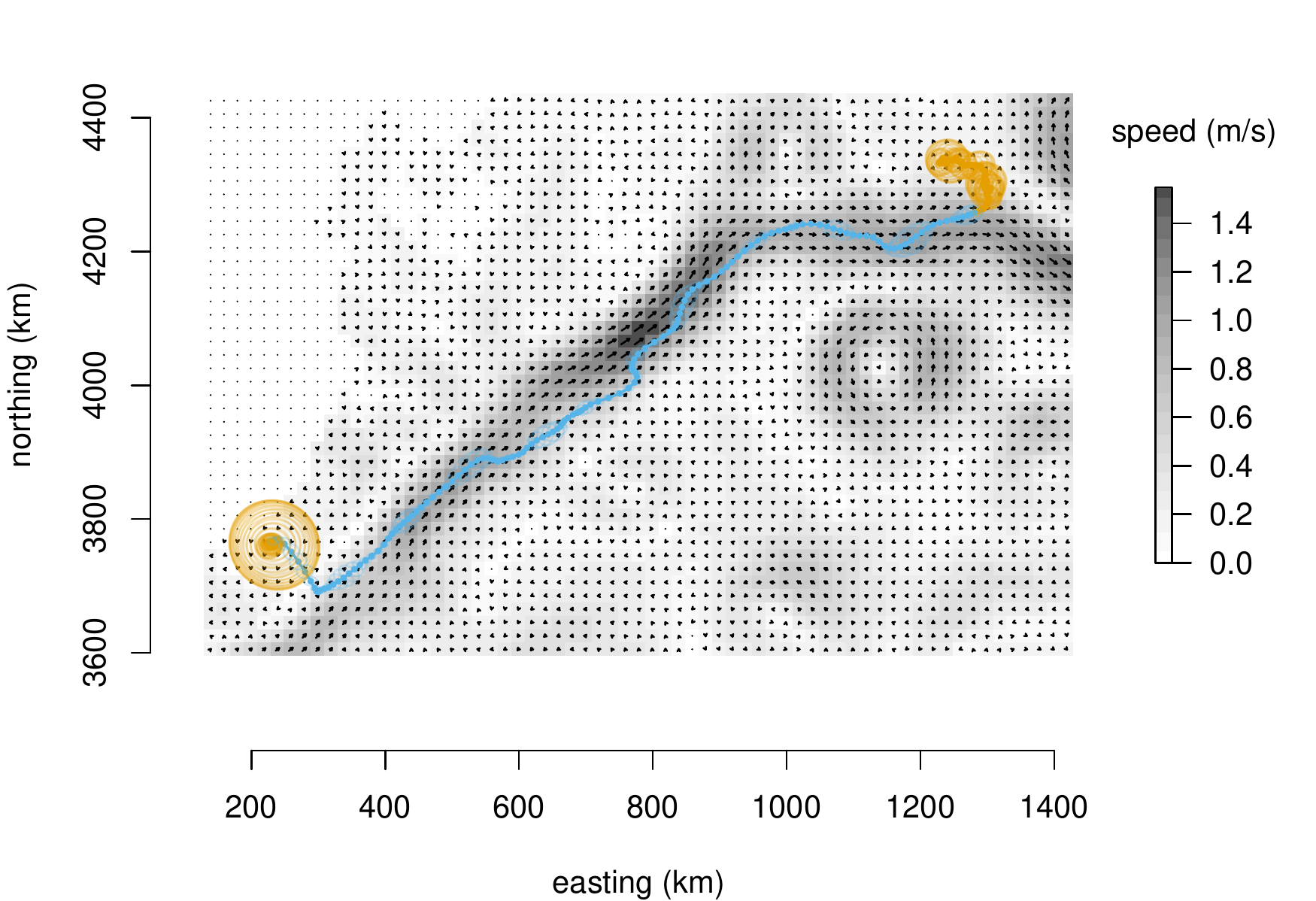}
  %\end{adjustbox}
  \caption{Selected results from the loggerhead turtle example. Top panels include estimates and 95\% confidence intervals for the ``transit'' state mean step length parameter as a function of ocean surface current speed $(w)$ when ocean surface current direction $(r_t)$ is the same as the bearing $(b_t)$ of movement (i.e.\ ${\text angle\_osc}=\cos(b_t-r_t)=1$; top-left panel), mean step length parameter as a function of ${\text angle\_osc}$ at the mean ocean surface current speed ($w=0.46$ m/s; top-middle panel), and mean turning angle parameter as a function of $d_t=\text{atan2}(\sin(r_t-b_{t-1}),\cos(r_t-b_{t-1}))$ (top-right panel). Bottom panel plots the pooled track, 95\% error ellipse confidence band, and state (orange = ``foraging'', blue = ``transit'') estimates based on multiple imputations of the position process relative to ocean surface current speed (m/s) and direction on 2 December 2012. Estimates therefore reflect uncertainty attributable to location measurement error and temporally-irregular observations. The turtle spent 0.57 (95\% CI: 0.48$-$0.65) of the 2 h time steps in the ``foraging'' state and 0.43 (95\% CI: 0.35$-$0.52) of time steps in the ``transit'' state as it travelled northeast along a predominant current until it (presumably) found an attractive foraging patch.}
  \label{fig:turtleResults}
\end{figure}

Using multiple imputation, we fit a 2-state HMM including a ``foraging'' state unaffected by currents and a ``transit'' state potentially influenced by ocean surface currents% as in Eqs. \ref{eq:turtleMeanStep} and \ref{eq:turtleMeanAngle}. We first used \verb|crawlWrap| to predict $T=350$ temporally-regular locations at 2 h time steps assuming a bivariate normal measurement error model that accounts for the Argos location quality class of each observation. We then used multiple imputation to account for location uncertainty by repeatedly fitting the HMM to $m=100$ realizations of the position process using \verb|MIfitHMM|
. After preparing the data, this rather complicated HMM can be specified, fitted, and visualized in only a few lines of code (see package vignette for complete details). %:
%<<fit-turtle-2, echo=TRUE, eval=FALSE>>=
%nbStates <- 2
%dist <- list(step = "gamma", angle = "wrpcauchy")
%DM <- list(step = list(mean = ~state2(w:angle_osc), sd = ~1),
%           angle = list(mean = ~state2(d), concentration= ~1))
%turtleFits <- MIfitHMM(miData, nbStates = nbStates, dist = dist, 
%                       Par0 = Par0, DM = DM, 
%                       estAngleMean = list(angle = TRUE),
%                       circularAngleMean = list(angle = TRUE))
%plot(turtleFits, plotCI = TRUE, covs = data.frame(angle_osc = cos(0)))
%@
%\noindent where \verb|angle_osc| $=\cos(b_t-r_t)$ and \verb|d| $=\text{atan2}(\sin(r_t-b_{t-1}),\cos(r_t-b_{t-1}))$.%Note that the \verb|state2| special function in \verb|DM| indicates the covariate formulas are specific to state 2 (``transit''), $\text{angle\_osc} = \cos(b_t-r_t)$, and the \verb|circularAngleMean| argument indicates that the circular-circular regression link function is to be used on the mean turning angle parameter as in Eq.\ \ref{eq:turtleMeanAngle}. Complete syntax details and reproducible code for this example can be found in the package vignette and supporting materials.
For the ``transit'' state, pooled parameter estimates indicated step lengths increased with ocean surface current speed and as the bearing of movement aligned with ocean surface current direction (%$\beta^l_1=0.39, \text{95\% CI: } 0.12-0.65$; 
Fig.\ \ref{fig:turtleResults}). The estimated wrapped Cauchy distribution for turning angle had mean angles $(\mu_t)$ 
biased towards the direction of ocean surface currents for each time step%$(\beta^\phi=0.26, \text{95\% CI: } 0.01-0.51)$
, with concentration parameter $\rho=0.86$ (95\% CI: 0.8$-$0.92) indicating turning angles were highly concentrated around $\mu_t$. Thus movement during the ``transit'' state appears to strongly follow ocean surface currents% (mean $\text{angle\_osc}=0.87,\text{SD}=0.22$), 
, while movement during the ``foraging'' state exhibited shorter step lengths %$(\mu^l_1=3066{\text m}, \text{95\% CI: } 2148-3984)$ 
perpendicular to ocean surface currents %(mean $\text{angle\_osc}=0.07,\text{SD}=0.27$)
with no directional persistence% $(\rho^\phi_1=0.48)$
. %The turtle spent 0.57 (95\% CI: 0.48$-$0.65) of the 2 h time steps in the ``foraging'' state and 0.43 (95\% CI: 0.35$-$0.52) of time steps in the ``transit'' state as it travelled northeast along a predominant current until it (presumably) found an attractive foraging patch (Fig.\ \ref{fig:turtleResults}).

\subsection{Grey seal}
\label{sec:greySeal}
We now revisit an analysis of a grey seal ({\it Halichoerus grypus}) track that was originally conducted by \cite{McClintockEtAl2012} using Bayesian methods and computationally-intensive Markov chain Monte Carlo. This seal tended to move in a clockwise fashion between two terrestrial ``haul-out'' sites and a foraging area in the North Sea. %The data consist of 1045 temporally-irregular Fastloc GPS locations collected in the North Sea between 9 April and 11 August 2008. 
Because the seal repeatedly visited the same haul-out and foraging locations, it provides an excellent example for demonstrating how biased movements relative to activity centres can be modelled using \verb|momentuHMM|. \cite{McClintockEtAl2012} fitted a 5-state HMM to these data that included three ``centre of attraction'' states, with movement biased towards two haul-out sites (``Abertay'' and ``Farne Islands'') and a foraging area (``Dogger Bank''), and two ``exploratory'' states %(''low speed'', ''high speed'') 
that were unassociated with an activity centre. After using \verb|crawlWrap| to predict $T=1515$ temporally-regular locations at 2 h time steps, we performed a very similar analysis to \cite{McClintockEtAl2012} using \verb|MIfitHMM|%by modelling the movement and transition probability parameters as a function of distance and angle to activity centres in a fashion similar to section \ref{sec:turtle}. See the package vignette for further details and code
.

\begin{figure}[htbp]
  \centering
  %\begin{adjustbox}{trim=0cm 0.25cm 0cm 1.5cm}
    \includegraphics[width=0.8\textwidth]{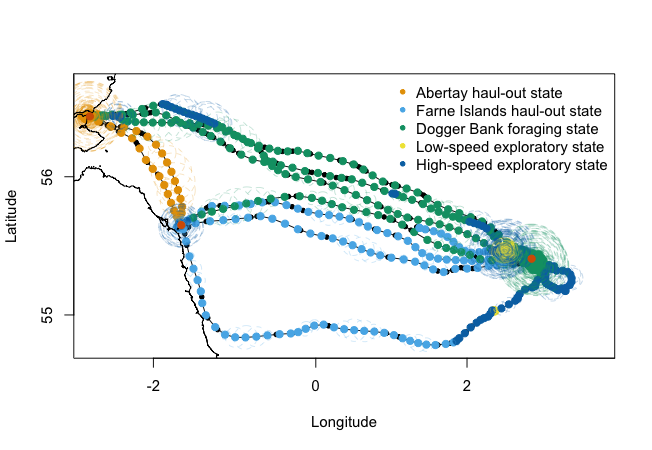}\\
    \includegraphics[width=0.8\textwidth]{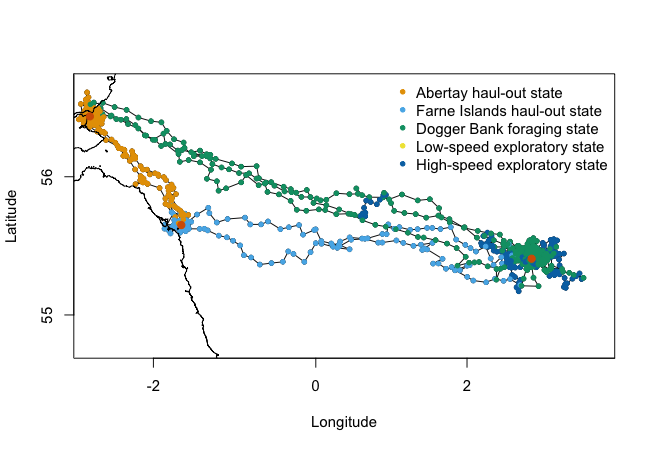}
  %\end{adjustbox}
  \caption{Fitted and simulated tracks from the grey seal example. This seal tended to move in a clockwise fashion between two haul-out locations (``Abertay'' and ``Farne Islands'') and a foraging area (``Dogger Bank'') in the North Sea. Top panel plots the pooled track, 95\% error ellipse confidence bands, and state estimates based on the 5-state HMM fitted to multiple imputations of the position process. Red points indicate the locations of the three activity centres. Black points indicate the (temporally-irregular) observed locations. Bottom panel presents the locations and states for a track simulated from the fitted model using the `simData' function.}
  \label{fig:greySealStateSims}
\end{figure}

%As in \cite{McClintockEtAl2012}, we assumed a Weibull distribution for step length where both the shape and scale parameters depended on the distance from the location at time $t$ to each activity centre. For the activity centres on land (``Abertay'' and ``Farne''), we allowed the (state-dependent) step length parameters to change when the seal was beyond 2500m of the haulout. For the ``Dogger'' activity centre, we allowed the parameters to change when the seal was beyond 15000m of this (presumably) foraging area. We thus allowed the movement behaviour to change within these activity centre states upon entering or leaving the vicinity of these sites.  We assumed a wrapped Cauchy distribution for turning angle with (state-dependent) mean angle derived from the direction to each activity centre at time $t$, and the concentration parameter was modelled similarly to the step length parameters. For the two ``exploratory'' states, we assumed they were simple random walks unaffected by proximity to activity centres. The state transition probabilities were also allowed to change as a function of distance to activity centres. To complete our model specification, we used the \verb|knownStates| argument to assign the seal to the corresponding activity centre state whenever it was within the 2500m (haulout area) or 15000m (foraging area) thresholds for each imputed data set, and the \verb|fixPar| argument was used to remove short-term directional persistence (and thus formulate the model as a mixture of biased and simple random walks). 
Each model fit required about 4 min on a standard desktop computer (macOS El Capitan, 16 GB RAM, 2.8 GHz Intel Core i7). Estimated activity budgets for the 5 states of this multiple-imputation HMM were 0.28 $(0.28-0.29)$ for the ``Abertay'' haul-out state, 0.12 $(0.11-0.14)$ for the ``Farne Islands'' haul-out state, 0.37 $(0.35-0.38)$ for the ``Dogger Bank'' foraging state, 0.09 $(0.04-0.18)$ for a low-speed ``exploratory'' state, and 0.14 $(0.09-0.22)$ for a high-speed ``exploratory'' state. All three activity centre states exhibited shorter step lengths and less biased movements when within the vicinity of these targets. Results from this analysis were thus very similar to \cite{McClintockEtAl2012}, but this implementation required far less computation time and no custom model-fitting algorithms. A simulated track generated using \verb|simData| is presented along with the fitted track in Fig.\ \ref{fig:greySealStateSims}.% While potentially useful for study design, power analysis, and prediction, the \verb|simData| function can also be helpful in assessing goodness of fit by repeatedly drawing simulated data sets from a fitted model and comparing them to observed properties of the data \citep[e.g.][]{MoralesEtAl2004}.

\section{Discussion}
\label{sec:discuss}
We have introduced the R package \verb|momentuHMM| and demonstrated some of its capabilities for conducting multivariate HMM analyses with animal location, auxiliary biotelemetry, and environmental data. %The package allows for fitting and simulating from a suite of biased and correlated random walk movement process models \citep[e.g.][]{McClintockEtAl2012}, can be used for an unlimited number of data streams and latent behaviour states, includes multiple imputation methods to account for observation error and missing data% that would otherwise be prohibitive to maximum likelihood analysis
%, and integrates seamlessly with rasters to facilitate spatio-temporal covariate modelling. 
There are many types of analyses that can be conducted with \verb|momentuHMM| that hitherto required custom code and model-fitting algorithms \citep[e.g.][]{McClintockEtAl2013c,McClintockEtAl2017,LangrockEtAl2014,MichelotEtAl2017}%, including the 3-state harbour seal model of \cite{McClintockEtAl2013c}, the 4-state southern elephant seal model of \citeauthor{MichelotEtAl2017} (\citeyear{MichelotEtAl2017}; Fig. \ref{fig:sesTracks}), the 6-state bearded seal model of \cite{McClintockEtAl2017}, and the group dynamic model of \cite{LangrockEtAl2014}
. The package therefore greatly expands on available software and facilitates the incorporation of more ecological and behavioural realism for hypothesis-driven analyses of animal movement that account for many of the challenges commonly associated with telemetry data. While many features were motivated by animal movement data, we note that the package can be used for analyzing any type of data that is amenable to HMMs.

Model fitting is relatively fast because the forward algorithm (Eq.\ \ref{eq:HMMlike}) is coded in C++. Because multiple imputations are completely parallelizable, with sufficient processing power computation times for analyses that account for measurement error, temporal irregularity, or other forms of missing data need not be longer than that required to fit a single HMM.  However, computation times will necessarily be longer as the number of states and/or parameters increase. For example, on a standard desktop computer (macOS El Capitan, 16 GB RAM, 2.8 GHz Intel Core i7), \verb|momentuHMM| required about 1 hour to fit a single HMM with $N=6$ states, seven data streams, and $T=7414$ time steps \citep{McClintock2017}.

As in any maximum likelihood analysis based on numerical optimization, computation times will also depend on starting values. Specifying ``good'' starting values is arguably the most challenging aspect of model fitting, and the package includes functions that are designed to help with the specification of starting values (e.g. \verb|checkPar0|, \verb|getPar|, \verb|getPar0|, and \verb|getParDM|). It also includes options for re-optimization based on random perturbations of the parameters to help explore the likelihood surface and diagnose convergence to local maxima. %Optimization for the circular-linear regression link function (\verb|tan(mean/2)|; see Table \ref{tab:pdfs}) in particular can be prone to local minima, so users are encouraged to explore a range of starting values when fitting these models.

While \verb|momentuHMM| includes functions for drawing realizations of the position process based on the CTCRW model of \cite{JohnsonEtAl2008}, this is but one of many methods for performing multiple imputation. Realizations of the position process from any movement model that accounts for measurement error and/or temporal irregularity \citep[e.g.][]{CalabreseEtAl2016} can be passed to \verb|MIfitHMM|. Multiple imputation methods also need not be limited to these telemetry error scenarios. For example, other forms of missing data could be imputed, thereby allowing investigation of non-random mechanisms for missingness that can be problematic if left unaccounted for in HMMs.

There remain many potential avenues for refinement and extension. Computation times could likely be improved by further optimizing the code for speed. Notable extensions include hidden semi-Markov models and random effects %on data stream probability distribution and state transition probabilitiy parameters 
\citep{ZucchiniEtAl2016}. %We would also like to incorporate additional parameters for change-point thresholds and the locations of activity centres instead of requiring that they be pre-specified (and potentially compared using AIC or other model selection criteria) as in grey seal example.
It is straightforward to add additional data stream probability distributions, and practitioners interested in additional features are encouraged to contact the authors.

\section*{Acknowledgments}
We are grateful to R. Scott, B. Godley, M. Godfrey, J. Sudre, and North Carolina Aquariums for providing the data used in our turtle example. The findings and conclusions in the manuscript are those of the author(s) and do not necessarily represent the views of the National Marine Fisheries Service, NOAA. Any use of trade, product, or firm names does not imply an endorsement by the US Government.

\section*{Authors' contributions}
B.T.M. and T.M. developed the package and wrote the manuscript.

\section*{Data accessibility}
All code mentioned here is available in the \verb|momentuHMM| package for R  hosted  on  CRAN  at  \url{https://CRAN.R-project.org/package=momentuHMM}. The development version of the package is available on GitHub at \url{https://github.com/bmcclintock/momentuHMM}.

\bibliographystyle{mee}
\bibliography{master}

\begin{thebibliography}{25}
\providecommand{\natexlab}[1]{#1}

\bibitem[{Beyer \emph{et~al.}(2013)Beyer, Morales, Murray \&
  Fortin}]{BeyerEtAl2013}
Beyer, H.L., Morales, J.M., Murray, D. \& Fortin, M.J. (2013) The effectiveness
  of {B}ayesian state-space models for estimating behavioural states from
  movement paths.
\newblock \emph{Methods in Ecology and Evolution}, \textbf{4}, 433--441.

\bibitem[{Calabrese \emph{et~al.}(2016)Calabrese, Fleming \&
  Gurarie}]{CalabreseEtAl2016}
Calabrese, J.M., Fleming, C.H. \& Gurarie, E. (2016) ctmm: an {R} package for
  analyzing animal relocation data as a continuous-time stochastic process.
\newblock \emph{Methods in Ecology and Evolution}, \textbf{7}, 1124--1132.

\bibitem[{Costa \emph{et~al.}(2010)Costa, Robinson, Arnould, Harrison, Simmons,
  Hassrick, Hoskins, Kirkman, Oosthuizen, Villegas-Amtmann
  \emph{et~al.}}]{CostaEtAl2010}
Costa, D.P., Robinson, P.W., Arnould, J.P., Harrison, A.L., Simmons, S.E.,
  Hassrick, J.L., Hoskins, A.J., Kirkman, S.P., Oosthuizen, H.,
  Villegas-Amtmann, S. \emph{et~al.} (2010) Accuracy of {ARGOS} locations of
  pinnipeds at-sea estimated using {F}astloc {GPS}.
\newblock \emph{PloS one}, \textbf{5}, e8677.

\bibitem[{DeRuiter \emph{et~al.}(2017)DeRuiter, Langrock, Skirbutas, Goldbogen,
  Calambokidis, Friedlaender \& Southall}]{DeRuiterEtAl2017}
DeRuiter, S.L., Langrock, R., Skirbutas, T., Goldbogen, J.A., Calambokidis, J.,
  Friedlaender, A.S. \& Southall, B.L. (2017) A multivariate mixed hidden
  {M}arkov model to analyze blue whale diving behaviour during controlled sound
  exposures.
\newblock \emph{The Annals of Applied Statistics}, \textbf{11}, 362--392.

\bibitem[{Hijmans(2016)}]{Hijmans2016}
Hijmans, R.J. (2016) \emph{raster: Geographic Data Analysis and Modeling}.
\newblock R package version 2.5-8.

\bibitem[{Hooten \emph{et~al.}(2017)Hooten, Johnson, McClintock \&
  Morales}]{HootenEtAl2017}
Hooten, M.B., Johnson, D.S., McClintock, B.T. \& Morales, J.M. (2017)
  \emph{Animal Movement: Statistical Models for Telemetry Data}.
\newblock CRC Press.

\bibitem[{Johnson(2017)}]{Johnson2017}
Johnson, D.S. (2017) \emph{crawl: Fit Continuous-Time Correlated Random Walk
  Models to Animal Movement Data}.
\newblock R package version 2.1.1.

\bibitem[{Johnson \emph{et~al.}(2008)Johnson, London, Lea \&
  Durban}]{JohnsonEtAl2008}
Johnson, D.S., London, J.M., Lea, M.A. \& Durban, J.W. (2008) Continuous-time
  correlated random walk model for animal telemetry data.
\newblock \emph{Ecology}, \textbf{89}, 1208--1215.

\bibitem[{Jonsen \emph{et~al.}(2005)Jonsen, Flemming \& Myers}]{JonsenEtAl2005}
Jonsen, I.D., Flemming, J.M. \& Myers, R.A. (2005) Robust state--space modeling
  of animal movement data.
\newblock \emph{Ecology}, \textbf{86}, 2874--2880.

\bibitem[{Langrock \emph{et~al.}(2014)Langrock, Hopcraft, Blackwell, Goodall,
  King, Niu, Patterson, Pedersen, Skarin \& Schick}]{LangrockEtAl2014}
Langrock, R., Hopcraft, G., Blackwell, P., Goodall, V., King, R., Niu, M.,
  Patterson, T., Pedersen, M., Skarin, A. \& Schick, R. (2014) Modelling group
  dynamic animal movement.
\newblock \emph{Methods in Ecology and Evolution}, \textbf{5}, 190--199.

\bibitem[{Langrock \emph{et~al.}(2012)Langrock, King, Matthiopoulos, Thomas,
  Fortin \& Morales}]{LangrockEtAl2012}
Langrock, R., King, R., Matthiopoulos, J., Thomas, L., Fortin, D. \& Morales,
  J.M. (2012) Flexible and practical modeling of animal telemetry data: hidden
  {M}arkov models and extensions.
\newblock \emph{Ecology}, \textbf{93}, 2336--2342.

\bibitem[{McClintock(2017)}]{McClintock2017}
McClintock, B.T. (2017) Incorporating telemetry error into hidden markov models
  of animal movement using multiple imputation.
\newblock \emph{Journal of Agricultural, Biological, and Environmental
  Statistics}, \textbf{22}, 249--269.

\bibitem[{McClintock \emph{et~al.}(2014)McClintock, Johnson, Hooten, Ver~Hoef
  \& Morales}]{McClintockEtAl2014b}
McClintock, B.T., Johnson, D.S., Hooten, M.B., Ver~Hoef, J.M. \& Morales, J.M.
  (2014) When to be discrete: the importance of time formulation in
  understanding animal movement.
\newblock \emph{Movement Ecology}, \textbf{2}, 21.

\bibitem[{McClintock \emph{et~al.}(2012)McClintock, King, Thomas,
  Matthiopoulos, McConnell \& Morales}]{McClintockEtAl2012}
McClintock, B.T., King, R., Thomas, L., Matthiopoulos, J., McConnell, B.J. \&
  Morales, J.M. (2012) A general discrete-time modeling framework for animal
  movement using multistate random walks.
\newblock \emph{Ecological Monographs}, \textbf{82}, 335--349.

\bibitem[{McClintock \emph{et~al.}(2017)McClintock, London, Cameron \&
  Boveng}]{McClintockEtAl2017}
McClintock, B.T., London, J.M., Cameron, M.F. \& Boveng, P.L. (2017) Bridging
  the gaps in animal movement: hidden behaviors and ecological relationships
  revealed by integrated data streams.
\newblock \emph{Ecosphere}, \textbf{8}, e01751.

\bibitem[{McClintock \emph{et~al.}(2013)McClintock, Russell, Matthiopoulos \&
  King}]{McClintockEtAl2013c}
McClintock, B.T., Russell, D.J., Matthiopoulos, J. \& King, R. (2013) Combining
  individual animal movement and ancillary biotelemetry data to investigate
  population-level activity budgets.
\newblock \emph{Ecology}, \textbf{94}, 838--849.

\bibitem[{McCullagh \& Nelder(1989)}]{McCullaghNelder1989}
McCullagh, P. \& Nelder, J.A. (1989) \emph{Generalized Linear Models, Second
  Edition}.
\newblock Chapman and Hall, New York.

\bibitem[{Michelot \emph{et~al.}(2017)Michelot, Langrock, Bestley, Jonsen,
  Photopoulou \& Patterson}]{MichelotEtAl2017}
Michelot, T., Langrock, R., Bestley, S., Jonsen, I.D., Photopoulou, T. \&
  Patterson, T.A. (2017) Estimation and simulation of foraging trips in
  land-based marine predators.
\newblock \emph{Ecology}, \textbf{98}, 1932--1944.

\bibitem[{Michelot \emph{et~al.}(2016)Michelot, Langrock \&
  Patterson}]{MichelotEtAl2016}
Michelot, T., Langrock, R. \& Patterson, T.A. (2016) move{HMM}: An {R} package
  for the statistical modelling of animal movement data using hidden {M}arkov
  models.
\newblock \emph{Methods in Ecology and Evolution}, \textbf{7}, 1308--1315.

\bibitem[{Morales \emph{et~al.}(2004)Morales, Haydon, Frair, Holsinger \&
  Fryxell}]{MoralesEtAl2004}
Morales, J.M., Haydon, D.T., Frair, J., Holsinger, K.E. \& Fryxell, J.M. (2004)
  Extracting more out of relocation data: building movement models as mixtures
  of random walks.
\newblock \emph{Ecology}, \textbf{85}, 2436--2445.

\bibitem[{{R Core Team}(2017)}]{RCoreTeam2017}
{R Core Team} (2017) \emph{R: A Language and Environment for Statistical
  Computing}.
\newblock R Foundation for Statistical Computing, Vienna, Austria.

\bibitem[{Rivest \emph{et~al.}(2016)Rivest, Duchesne, Nicosia \&
  Fortin}]{RivestEtAl2016}
Rivest, L.P., Duchesne, T., Nicosia, A. \& Fortin, D. (2016) A general angular
  regression model for the analysis of data on animal movement in ecology.
\newblock \emph{Journal of the Royal Statistical Society: Series C (Applied
  Statistics)}, \textbf{65}, 445--463.

\bibitem[{Wall \emph{et~al.}(2014)Wall, Wittemyer, LeMay, Douglas-Hamilton \&
  Klinkenberg}]{WallEtAl2014}
Wall, J., Wittemyer, G., LeMay, V., Douglas-Hamilton, I. \& Klinkenberg, B.
  (2014) Elliptical time-density model to estimate wildlife utilization
  distributions.
\newblock \emph{Methods in Ecology and Evolution}, \textbf{5}, 780--790.

\bibitem[{Whoriskey \emph{et~al.}(2017)Whoriskey, Auger-M{\'e}th{\'e},
  Albertsen, Whoriskey, Binder, Krueger \& Mills~Flemming}]{WhoriskeyEtAl2017}
Whoriskey, K., Auger-M{\'e}th{\'e}, M., Albertsen, C.M., Whoriskey, F.G.,
  Binder, T.R., Krueger, C.C. \& Mills~Flemming, J. (2017) A hidden markov
  movement model for rapidly identifying behavioral states from animal tracks.
\newblock \emph{Ecology and Evolution}, \textbf{7}, 2112--2121.

\bibitem[{Zucchini \emph{et~al.}(2016)Zucchini, MacDonald \&
  Langrock}]{ZucchiniEtAl2016}
Zucchini, W., MacDonald, I.L. \& Langrock, R. (2016) \emph{Hidden Markov Models
  for Time Series: An Introduction Using R}.
\newblock CRC Press.

\end{thebibliography}

\end{document}